\documentclass[9pt,twocolumn,twoside]{optica}

\usepackage{subcaption}
\usepackage{overpic}
\setboolean{shortarticle}{false} 
\usepackage{subeqnar,overpic}
\usepackage{ulem}

\newtheorem{proposition}{Proposition}
\newtheorem{lemma}{Lemma}

\DeclareMathOperator{\dn}{dn}
\DeclareMathOperator{\cn}{cn}
\DeclareMathOperator{\sn}{sn}
\DeclareMathOperator{\sech}{sech}
\let\Re\relax
\DeclareMathOperator{\Re}{Re}
\let\Im\relax
\DeclareMathOperator{\Im}{Im}
\DeclareMathOperator{\vecspan}{span}

\newcommand{\per}{{\textnormal{per}}}

\ifthenelse{\boolean{shortarticle}}{\colorlet{color2}{color2b}}{\colorlet{color2}{color2}} % Automatically switches colors for short articles

\title{Stability and Dynamics of Microring Combs:  Elliptic function solutions of the Lugiato-Lefever equation}

\author[1]{Chang Sun}
\author[2]{Travis Askham}
\author[1,2,*]{J. Nathan Kutz}

\affil[1]{Department of Physics, University of Washington, Seattle, WA 90195}
\affil[2]{Department of Applied Mathematics, University of Washington, Seattle, WA 90195-3925}

\affil[*]{Corresponding author: kutz@uw.edu}

\dates{Compiled \today}

\ociscodes{(140.4780)  Optical resonators; (140.3945)   Microcavities;  (060.5530) Pulse propagation and temporal solitons;  (190.0190)  Nonlinear optics}

\doi{\url{http://dx.doi.org/10.1364/josab.XX.XXXXXX}}

\begin{abstract}
We consider a new class of periodic solutions to the Lugiato-Lefever equations (LLE) that govern the electromagnetic field in a microresonator cavity.   Specifically, we rigorously characterize the stability and dynamics of the Jacobi elliptic function solutions of the LLE and show that the $\dn$ solution is stabilized by the pumping of the microresonator.  In analogy with soliton perturbation theory, we also derive a microcomb perturbation theory that allows one to consider the effects of physically realizable perturbations on the comb line stability, including effects of Raman scattering and stimulated emission.  Our results are verified through full numerical simulations of the LLE cavity dynamics.  The perturbation theory gives a simple analytic platform for potentially engineering new resonator designs.
\end{abstract}

\setboolean{displaycopyright}{true}

\begin{document}

\maketitle
\thispagestyle{fancy}

\ifthenelse{\boolean{shortarticle}}{\ifthenelse{\boolean{singlecolumn}}{\abscontentformatted}{\abscontent}}{}

\section{Introduction}

Frequency comb generation in microresonators has become a critically enabling technology for applications in metrology, high-resolution spectroscopy and microwave photonics~\cite{app2,kippenberg2011microresonator,del2011octave,papp2011spectral,ferdous2011spectral,app1,liang2015high,suh2016microresonator}.   A clear goal in such microresonators is the generation of octave-spanning combs, which is often achieved by the generation of a single soliton in a high-Q microresonator cavity~\cite{herr2014temporal,yi2015soliton}.   
Much like the multi-pulsing instability (MPI) in mode-locked laser cavities~\cite{kutz1,kutz2,kutz3}, microresonators are prone to generating multiple pulses in the cavity~\cite{kip1,kip2}, thus compromising the performance of the frequency comb generation.   Consequently, the dynamics and stability of pulse generation in the microresonator is of significant interest.  In this manuscript, we explore analytically tractable solutions of the Lugiato-Lefever equation (LLE)~\cite{lugiato1987spatial}, which is the governing equation for the microresonator dynamics~\cite{PhysRevA.87.053852}.  While solitons have been observed in a number of experimental architectures, the deterministic manipulation of states with multiple solitons in microresonators has only been recently explored with the goal of prediction and control~\cite{kip1}.  We develop a perturbation theory for periodic pulse train solutions, known as Jacobi elliptic functions, which characterize the underlying solutions in the microresonator cavity. Our work provides a theoretically rigorous complement to recent experimental observations for the transitions between $N$ to $N+1$ (or vice versa) pulses in a microresonator.   We further show how cavity perturbations, due to, for instance, the Raman effect or spontaneous emission noise, affect the resulting combline stability and robustness. 

Soliton perturbation theory has been one of the most successful theoretical tools developed for characterizing the underlying physics in optical communication systems~\cite{karpman1977perturbation,kodama1981perturbations,kaup1990perturbation,elgin1993perturbations} and mode-locked lasers~\cite{kartner1996soliton,kapitula2002stability,kapitula2004evans,Bale:08}.  In this work, we develop a LLE combline perturbation theory.  The theory relies on an analytic solution, the Galilean invariant one-soliton solution, of the nonlinear Schr\"odinger equation.  Jacobi elliptic functions are a generalization of soliton solutions of the LLE equation, capable of representing both single localized pulse solutions and periodic pulse trains.    Much like solitons, the solutions are parameterized by a number of free parameters whose slow evolution under perturbation characterizes the stability of the solution.  A linear stability analysis of the Jacobi elliptic solutions is capable of revealing key properties of the combline properties under perturbation.   Specifically, our analysis characterizes the stability of $N$ pulses per round trip in the laser cavity.  Much like MPI in mode-locked lasers~\cite{kutz1,kutz2,kutz3}, an initial cavity cold start will jump to the most energetically favorable configuration.  However, our analysis shows how one can manipulate the number of pulses per round trip by simply manipulating the microresonator detuning, confirming experimental findings.

From a technical point of view, our stability analysis follows closely the rigorous theory of soliton perturbation theory. For the LLE {Jacobi elliptic solutions}, the linearized operator contains four zero modes which correspond to invariances of the solutions.  The effect of perturbations on these zero modes is quantified and shows how various perturbations can either destabilize the solutions or force solutions to translate or bifurcate to a higher or lower number of {pulses} per round trip.  Additionally, we show that the remainder of the continuous spectrum remains bounded under perturbation.  We demonstrate the application of the theory on two canonical examples: (i) the LLE modified to include Raman dynamics, and (ii) the LLE under the influence of white noise (stimulated emission noise) perturbations.  In both cases, we show that the comblines remain stable while the solitons undergo translation.  Given the tremendous impact that soliton perturbation theory has had on theoretical understanding of light-wave transmission systems, our goal is to provide a similar LLE combline perturbation theory for theoretical characterization of microresonators.

The paper is outlined as follows:  In Sec.~2, the LLE is introduced along with the scalings to be used in our perturbation theory.  Section~3 gives a brief overview of the perturbation theory to be used for modeling the microresonator.  Section~4 and 5 present the Jacobi elliptic function solutions that satisfy the LLE equation and their detailed linear stability analysis respectively.   The effects of two canonical perturbations due to Raman and simulated emission are considered in Sec. 6.  Section~7 provides a brief summary and outlook for the theoretical method developed.

\section{Lugiato Lefever equation}

The Lugiato-Lefever equation (LLE), which was originally derived in the context of detuned cavity resonators~\cite{lugiato1987spatial}, has been shown to describe the evolution of the electromagnetic field in microresonators~\cite{PhysRevA.87.053852}.  The LLE is a modification of the nonlinear Schr\"odinger equation (NLSE) which includes damping, detuning and a driving/pumping term.  In dimensionless form, the LLE is given by the
partial differential equation (PDE)
\begin{equation}
\frac{\partial u}{\partial t}=-(\epsilon+i\alpha)u+i|u|^{2}u-i\frac{\beta}{2}\frac{\partial ^{2}u}{\partial x^{2}}+\epsilon  F + \epsilon G(u,x,t),
\end{equation}
where $u(x,t)$ the complex envelope of the total intracavity electric field, $\beta$ determines the microring dispersion ($\beta>0$ is normal group-velocity dispersion while $\beta<0$ is anomalous group-velocity dispersion), $\alpha$ is the cavity detuning parameter, $F$ characterizes the external cavity pumping, and $x\in [-\pi,\pi)$ since the microresonator enforces periodic boundary conditions~\cite{PhysRevA.87.053852}.  In our specific scaling, the parameter $\epsilon\ll 1$ is used to model the effects of linear cavity attenuation and small perturbations of the form $G(u,x,t)$ to the dominant balance dynamics of dispersion, Kerr self-phase modulation, and detuning. 

In our scalings, the LLE can be written as a perturbed version of the detuned NLSE so that
\begin{equation}
i\frac{\partial u}{\partial t} - \frac{\beta}{2}\frac{\partial ^{2}u}{\partial x^{2}} +|u|^{2}u -\alpha u=i\epsilon(F-u + G(u,x,t)).
\label{eq:lle}
\end{equation}
This scaling allows us to develop a systematic perturbation analysis of previously unconsidered periodic, 
Jacobi elliptic solutions of the LLE.  This complements the  detailed stability analysis of Godey {\em et al}.~\cite{PhysRevA.89.063814}
which details the onset of a myriad of spatio-temporal patterns in the LLE model.   Specifically, they show that the steady-state
solutions of the LLE (with all temporal and spatial derivatives set to zero) lead to a host of pattern-forming instabilities~\cite{cross1993pattern} that
are ultimately responsible for the generation of strongly nonlinear periodic waveforms.    In our analysis, we consider the
stability of Jacobi elliptic solutions which are strongly nonlinear solutions whose dominant balance includes temporal and spatial derivative terms~\cite{carr2000stationary,bronski2001bose,bronski1,bronski2}.

\section{Background: Perturbation Theory}

Our stability analysis determines the spectrum of the resulting linearized operator along with the effects of perturbations on the evolution of the solution parameters.  In its most general form, we can consider the one dimensional PDE
\begin{equation}
   \frac{\partial u}{\partial t} = N(u, u_x, u_{xx}, \cdots, \mu) 
   +\epsilon  G(u,x,t) \; ,
\end{equation}
where $N(\cdot)$ represents some nonlinear dynamics (for which an 
analytical solution is known), $\epsilon G(u,x,t)$ 
is a perturbation to these dynamics, and $\mu$ is a (bifurcation) 
parameter. A multi-scale perturbation expansion~\cite{bender2013advanced,kevorkian2013perturbation} is a representation of 
the solution of the form
\begin{equation}
u(x,t)=u_{0}(x,t,\tau)+\epsilon u_{1}(x,t) + \epsilon^2 u_2 (x,t) + \cdots
\; ,
\label{eq:perturb}
\end{equation}
where $\tau=\epsilon t$ corresponds to a slow variable dependence~\cite{kodama1981perturbations,weinstein1985modulational}.

Collecting terms at each order of $\epsilon$ gives nonlinear dynamics
for the leading order term and forced, linear dynamics for
all other orders, i.e. 
\begin{subeqnarray}
 \frac{\partial u_0}{\partial t} &=& N (u_0, {u_0}_x, {u_0}_{xx}, \cdots, \mu)  \; ,\\
 \frac{\partial u_1}{\partial t} &=& L_1 (u_0) u_1 + F_1 (u_0)  \; ,\\
 \frac{\partial u_2}{\partial t} &=& L_2 (u_0) u_2 + F_2 (u_0,u_1) \; ,\\
  &\vdots& \nonumber
\end{subeqnarray}
where the first equation is the $O(1)$ balance, the second equation is the $O(\epsilon)$ balance
and the third equations is the $O(\epsilon^2)$ balance.
As in the approach of Weinstein \cite{weinstein1985modulational}, 
we consider a solution of the leading order problem with
slow-time modulation. Let $u_0(x,t)$ be given by
\begin{equation}
  u_0(x,t)= \Phi(x,t,A_1,A_2, \cdots) \; ,
\end{equation}
where the parameters $A_i(\tau)$ vary with 
the slow time scale $\tau$. 
Applying the Fredholm alternative 
to the forced, linear PDE for $u_1$ requires that the 
forcing term $F_1$ be orthogonal to the generalized
null space of the adjoint
operator $L_1^\dag$, i.e. if $(L_1^\dag)^m v =0$ for
some $m > 0$, then
\begin{equation}
  \langle v, F_1 \rangle = 0 \; ,
  \label{eq:fredholm} 
\end{equation}
where $\langle u,v \rangle=\int_D u v^* \, dx$ is the inner product 
over the domain $D$. For a given perturbation, this constraint
will result in equations for the slow evolution of the 
parameters $A_i$ of the form
\begin{equation}
  \frac{\partial {A_i}}{\partial \tau}  = f_i(A_1,A_2,\cdots) \; .
\end{equation}

Remarkably, in Weinstein's analysis of the NLSE \cite{weinstein1985modulational},
these constraints are all that needs to be satisfied to
show that $\epsilon u_1$ is small for small values of
$\epsilon$ up to times of order $1/\epsilon$.
Similar results hold for elliptic function solutions
of the NLSE, which we outline in the following. 
We will show that 
the additional terms in the LLE, when viewed as a 
perturbation of the NLSE, have a 
stabilizing effect on the parameters of $\dn$ type
solutions. Further, we provide expressions for the 
evolution of the parameters for two particular
cavity perturbations of the LLE: the Raman effect and 
spontaneous emission noise. 

\section{Jacobi Elliptic Functions for the NLSE}

The Jacobi elliptic functions are periodic wavefunctions that
satisfy the NLSE with
detuning~\cite{carr2000stationary,bronski2001bose,bronski1,bronski2},
i.e. the leading order dynamics as described by \eqref{eq:lle}.
The three basic functions are denoted $\sn(x|k)$, $\cn(x|k)$,
and $\dn(x|k)$, where the elliptic modulus, $k$,
parameterizes the solutions. The value of $k$ is constrained
such that $k \in [0,1)$; we note that the reader may be more
familiar with the parameter $m=k^2$, which is commonly used
in software for evaluating the Jacobi elliptic functions.

{
  The stability of these solutions is well-studied. For the
  defocusing case, the $\sn$ solutions are known to be
  modulationally stable \cite{Bernard}. For the focusing
  case, the $\cn$ and $\dn$ solutions are modulationally
  unstable \cite{deconinck2017stability}. Recent research
  has shown the spectral stability
  of the $\dn$ solution under perturbations with
  a period equal to the fundamental period, but not under
  perturbations with a period equal to a multiple of the
  fundamental period \cite{stabilityDn}.
  { Spectral stability of the $\cn(x|k)$ solution only holds when $k \in (0, k_{c})$ under perturbations with a period equal to the fundamental period, with $k_{c} \approx 0.908$ \cite{stabilityDn}.} In-depth discussion of
  the stability properties of Jacobi elliptic function solutions of
  the NLSE can be found in \cite{Bernard,deconinck2017stability,stabilityDn}.

  \begin{figure}[t]
    \includegraphics[width=\linewidth]{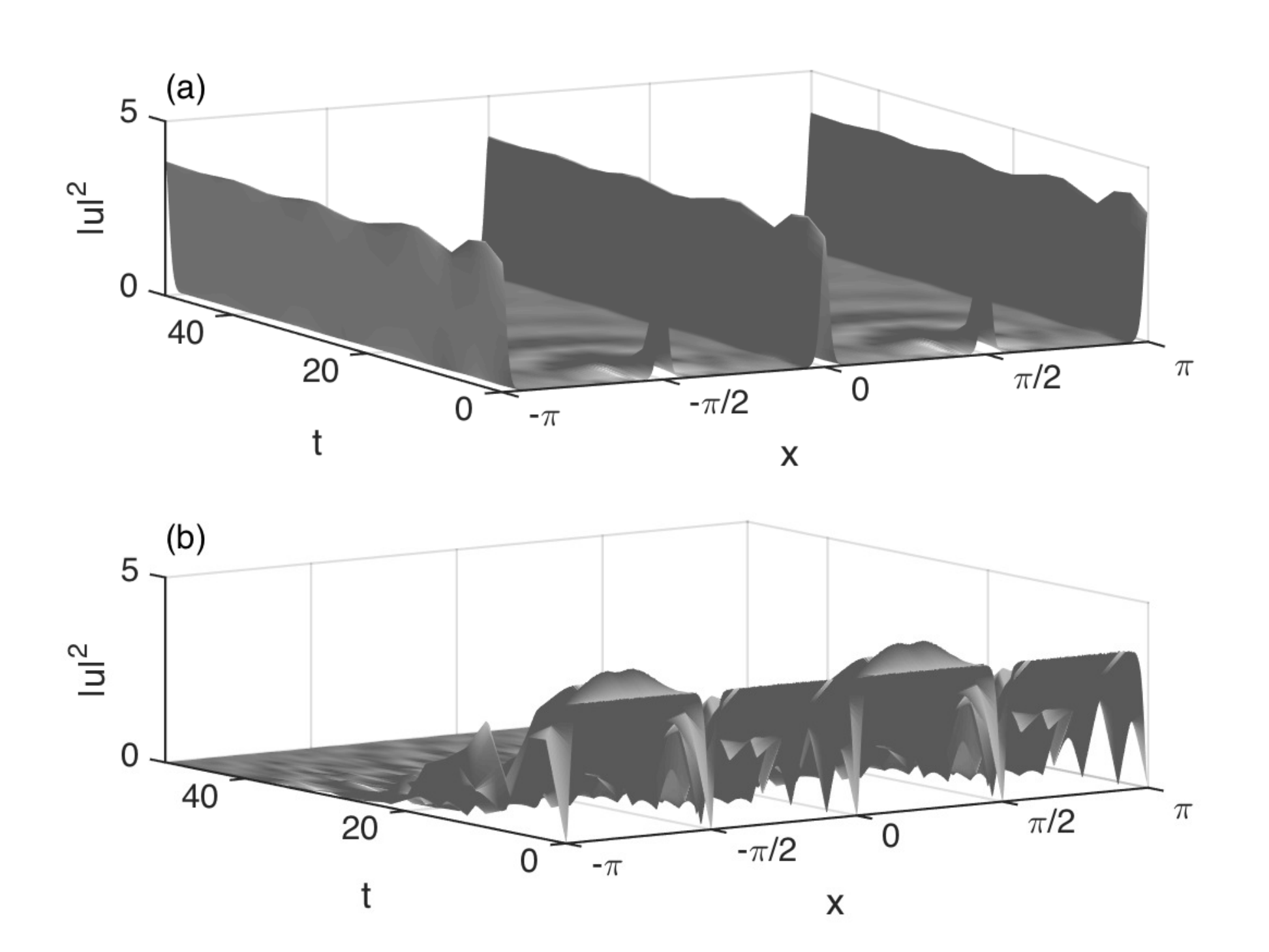}
    \caption{
      Numerical simulation of the (a) $\cn$ and (b) $\sn$ solutions { of
      \eqref{eq:lle} with $|\beta| = 0.01$, $\epsilon=0.1$, $G=0$, and the detuning $\alpha$ set to (a) $\alpha = 1.8732$ and (b) $\alpha = 3.7464$ (these values of the detuning are chosen so that $k^{2}=1-10^{-12}\approx 1$ in the analog of \eqref{eq:alpha} for these solutions).}  The solutions were seeded with a white noise perturbation to induce instability in the evolution.  Both solutions are unstable, even in the limit $k\rightarrow 1$ where the linear stability analysis shows the eigenvalues to shrink to the real axis.  Note that the $\cn$ solution collapses from an $N=4$ solution to a stable $N=2$ $\dn$ solution.}
    \label{fig:cn_sn}
  \end{figure}
  
  With the addition of the LLE terms, i.e. the damping and forcing
  of the microresonator, the $\cn$ and $\sn$ solutions are unstable
  in their respective regimes. In Figure~\ref{fig:cn_sn}, we plot
  a numerical simulation of the evolution of $\cn$ and $\sn$ wave
  forms (with four {pulses}) governed by the LLE. The
  $\sn$ wave form quickly decays and the $\cn$ wave form evolves
  into a solution of $\dn$ type (with two {pulses}). 
  It appears that the LLE does not support {pulses} that are separated
  by a node, i.e. those with a $\pi$ phase change between neighboring pulses.  
  In contrast
  with its instability as a solution of the NLSE, the $\dn$ type
  solutions of the LLE are in fact stable, even with
  multiple {pulses}
  in the cavity. We will show that this stability can be understood
  analytically and we will focus on the $\dn$ type solutions for
  the remainder of the manuscript.}

  %However, perturbations to the leading order dynamics can stabilize the periodic wavetrains.  In Bose-Einstein condensates, the addition of a Hamiltonian term, i.e. a periodic potential generated by the interference of lasers, can stabilize the elliptic function solutions\cite{bronski2001bose,bronski1,bronski2}.  In the LLE, the perturbations considered to \eqref{eq:lle} are dissipative in nature, modeling the damping and forcing on the microresonator.  As we will show, the dissipative terms can stabilize the $\dn(x|k)$ solutions in the anomalous regime, while the $\sn(x|k)$ and $\cn(x|k)$ solutions {with the dissipative terms are} unstable. Here we are more interested in the perturbed LLE, thus the following discussion are based on Jacobi elliptic function $\dn(x|k)$ since it's the only stabilized solution with the dissipative term on the right hand side.

\subsection*{Solutions of $\dn$ type:  anomalous dispersion}

The $\dn$ solution is of the most practical importance,
as it is the only stable solution we find for the LLE in
the anomalous dispersion regime {($\beta < 0$)}.
For this solution, we assume the general form
\begin{equation}
u_{0}(x,t)={{\hat u}_{0}}e^{i\psi}=A\dn(B(x-x_{0})|k)e^{i[\xi(x-x_{0})+\sigma-\sigma_{0}]}, 
\label{eq:dn}
\end{equation}
where $A^{2}=-\beta B^{2}$, and %{$\alpha = -{\beta}B^{2}(2-k^{2})/2$ This is not always true actually. This is only true when $d\sigma$/$dt$ and $\xi$ are 0. We can get rid of this here.}
\begin{subeqnarray}
&& \frac{dx_{0}}{dt}=-\beta\xi, \\
&& \frac{d\sigma}{dt}=-\alpha-\frac{\beta}{2}B^{2}(2-k^{2})-\frac{\beta}{2}\xi^{2}.
\end{subeqnarray}
Since the wavefunctions of the LLE should be ${2\pi}/{N}$ periodic, where $N$ is a positive integer, the value of $B$ determines the value of $k$ and vice-versa. Specifically, the period of the Jacobi elliptic function $y=\dn(x|k)$ is $2K$, where $K(k)$ is the elliptic integral of the first kind. So the period of $\hat{u}_{0}=\dn(Bx|k)$ should be $T={2K}/{B}$. If $T={2\pi}/{N}$, then we have ${2K}/{B}={2\pi}/{N}$, thus $B={KN}/{\pi}$.  Note that $N$ is the number of localized ({pulses}) per round trip in the microresonator.

\begin{figure}[t]
\begin{subfigure}{.23\textwidth}
  \centering
  \begin{overpic}[width=\linewidth]{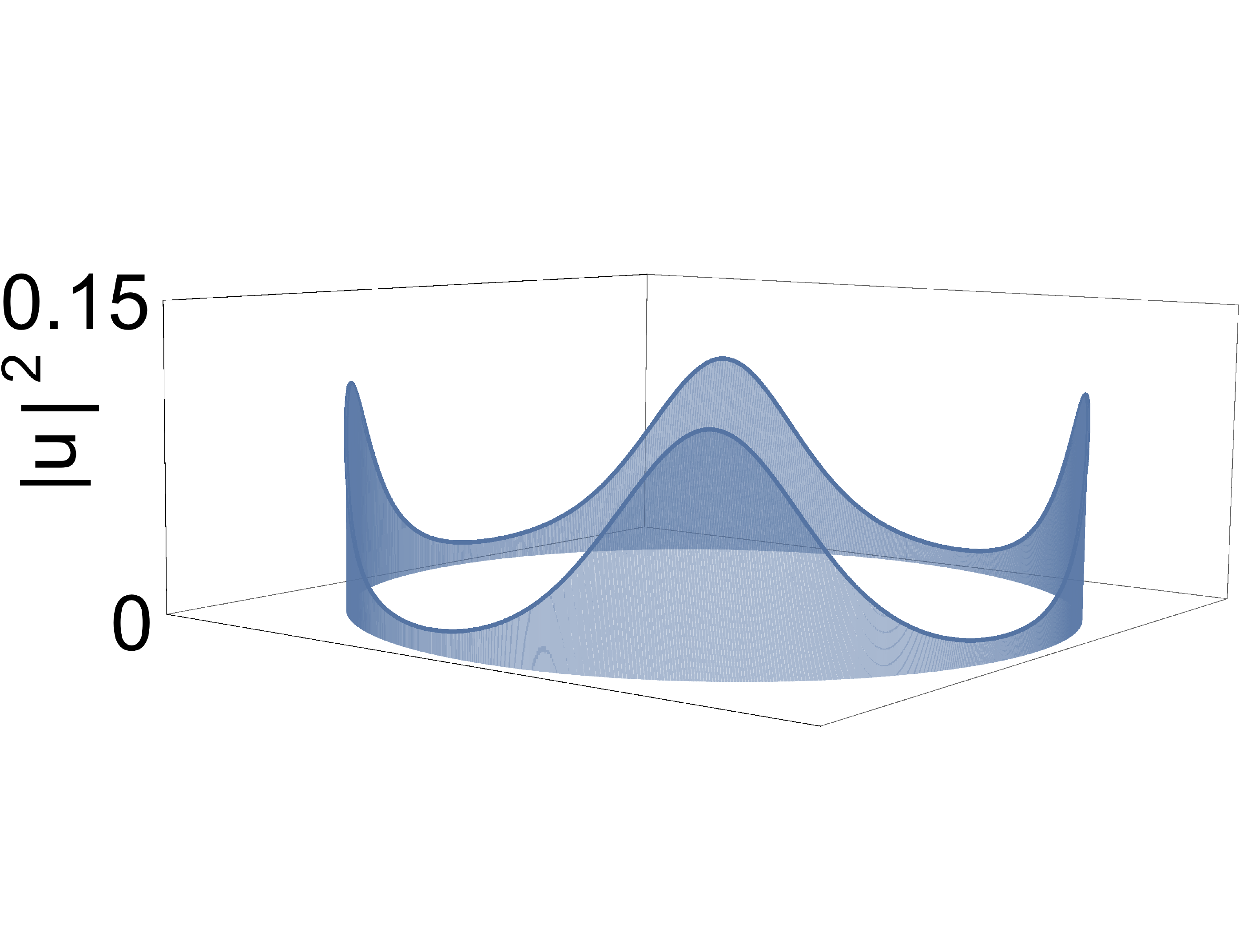}
  \put(17,14){(a)}
  \end{overpic}
 % \caption{$\psi = A\dn(B(x-x_{0}),k)e^{i(\xi(x-x_{0})+\sigma-\sigma_{0})}$, where $k^{2}=0.9$}
%  \label{fig:sfig1}
\end{subfigure}%
\begin{subfigure}{.23\textwidth}
  \centering
  \begin{overpic}[width=\linewidth]{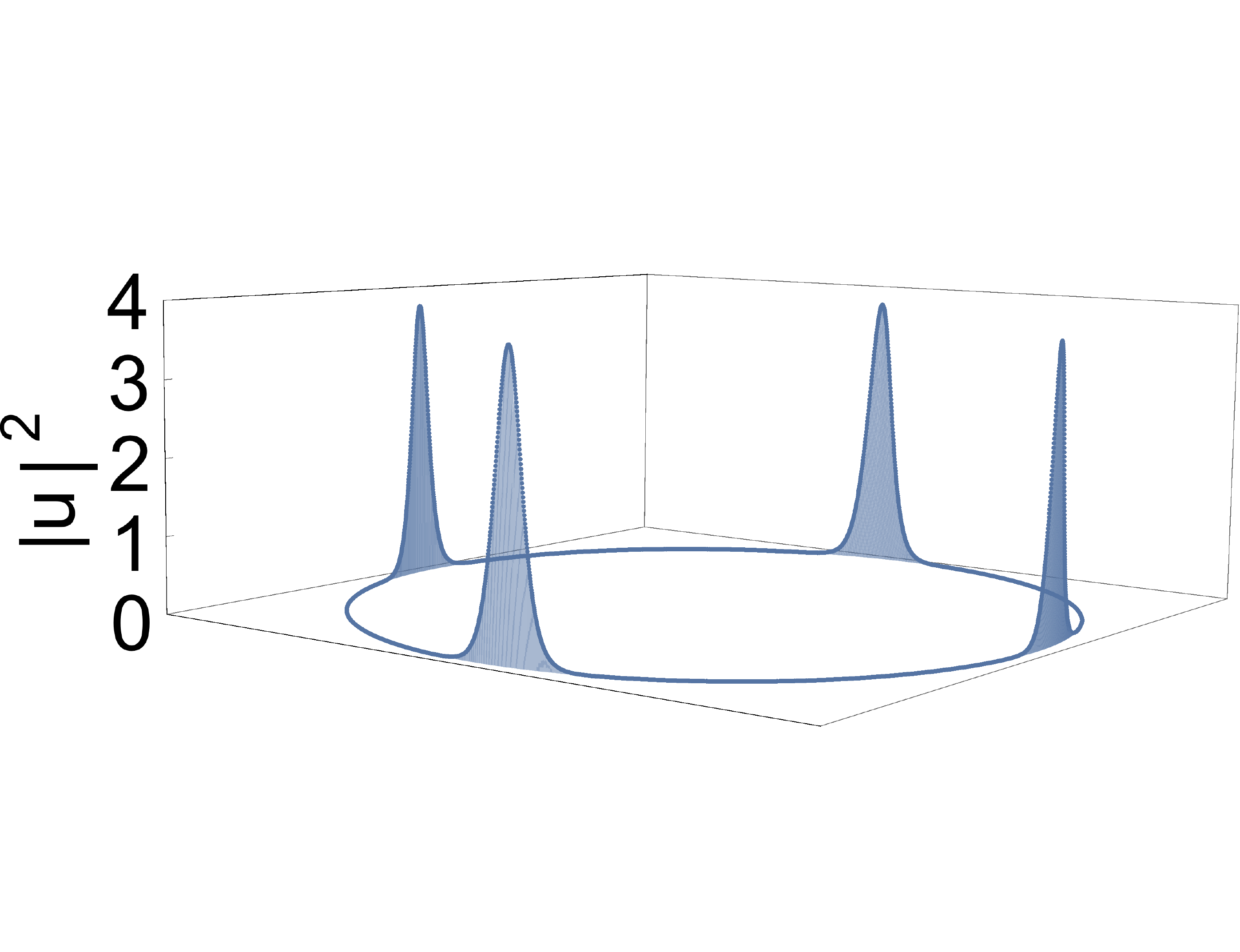}
  \put(17,14){(b)}
  \end{overpic}
 % \caption{$\psi = A\dn(B(x-x_{0}),k)e^{i(\xi(x-x_{0})+\sigma-\sigma_{0})}$, where $k^{2}=1-10^{-12}$}
%  \label{fig:sfig2}
\end{subfigure}
\vspace*{-0.2in}
\caption{The $\dn$-type solutions for (a) $k^2=0.9$ and (b) $k^2=1-10^{-12}\approx 1$ with $N=4$.  The two panels demonstrate that the elliptic modulus $k\in[0,1)$ can produce solutions which resemble a modulated CW beam or highly localized, hyperbolic secant pulses.  Note that the $\dn$ solution has no nodal points where the solution is zero.}
\label{fig:dn}
\end{figure}

Figure~\ref{fig:dn} shows the $\dn$ solution for two values of the parameter $k$, where $k \in [0,1)$.  These figures are illustrated with $N=4$ so that four {pulses} are shown around the cavity.  In the limit $k\rightarrow 1$, the function $\dn(x|k)  \rightarrow \sech(x)$, which is the standard hyperbolic secant soliton solution generated by the dominant NLSE terms.  When $k\rightarrow 0$, the function $\dn(x|k)\rightarrow 1$, which is a continuous wave solution of the LLE.  The figure illustrates the $k^2=0.9$ and $k^{2}=1-10^{-12}\approx 1$ solutions of the LLE.
\begin{figure}[t]
\begin{subfigure}{.23\textwidth}
  \centering
  \begin{overpic}[width=\linewidth]{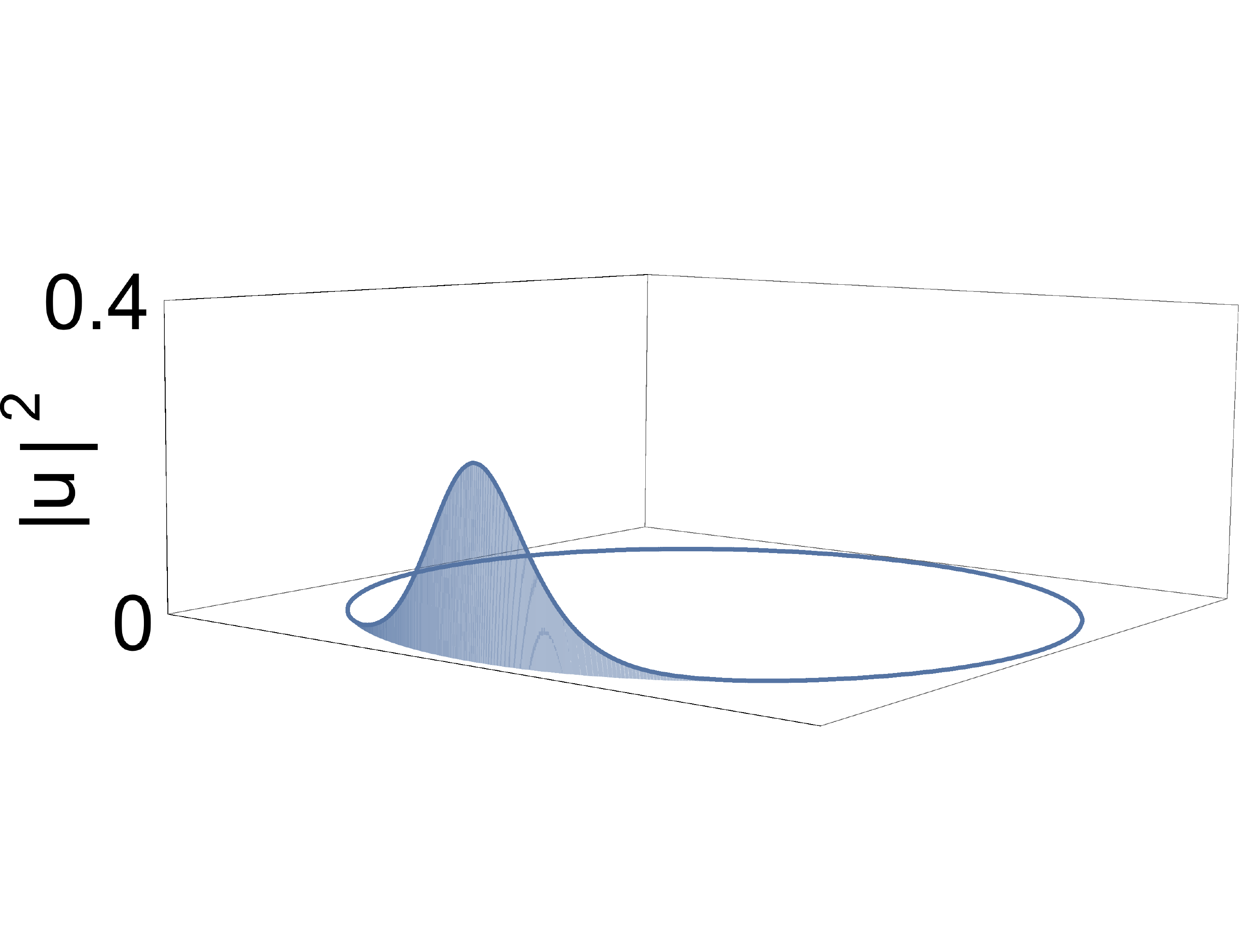}
  \end{overpic}
%  \caption{Number of solitons $n=1$}
%  \label{fig:sfig1}
\end{subfigure}%
\begin{subfigure}{.23\textwidth}
  \centering
  \begin{overpic}[width=\linewidth]{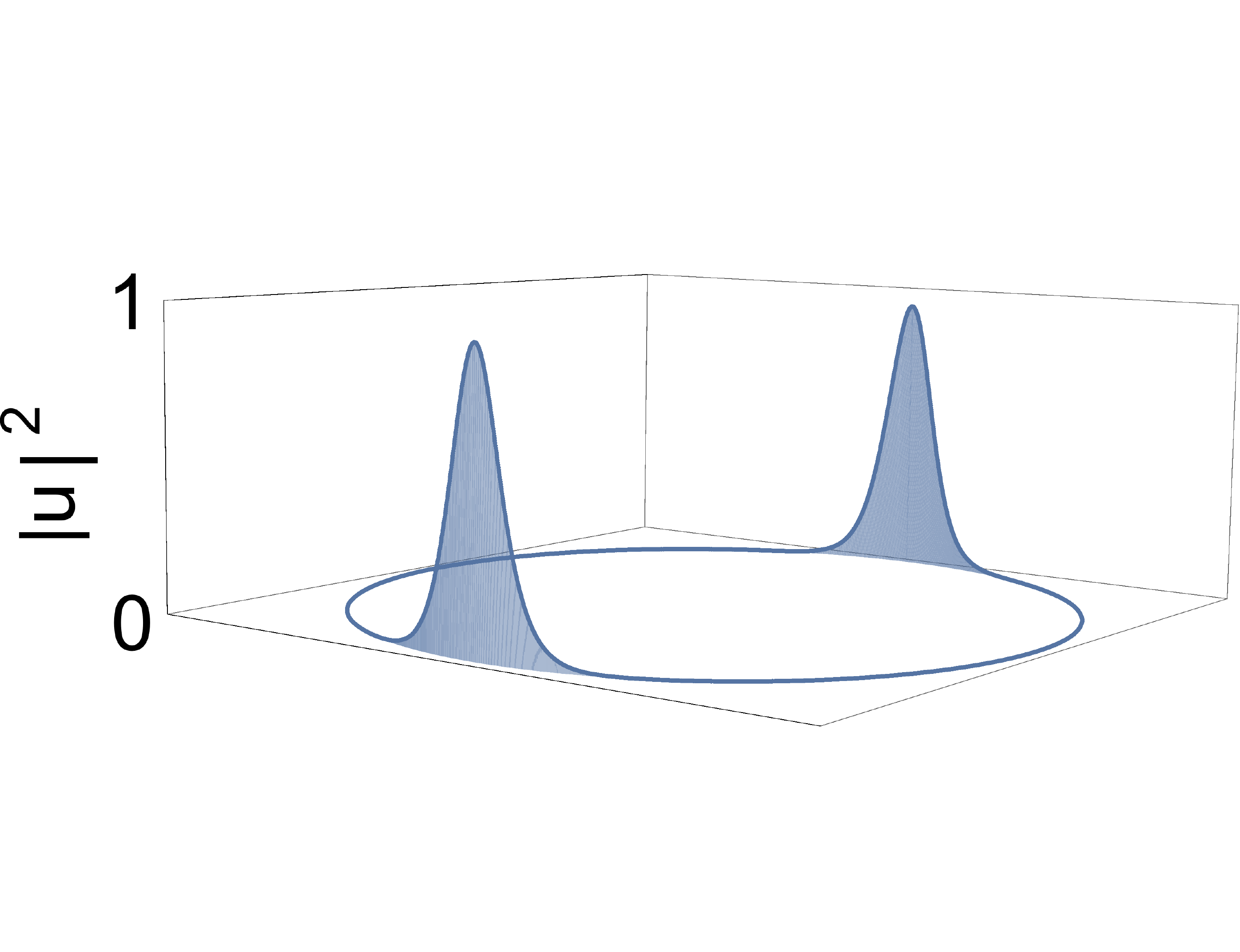}
  \end{overpic}
% \caption{Number of solitons $n=2$}
%  \label{fig:sfig2}
\end{subfigure}
\begin{subfigure}{.23\textwidth}
\vspace*{-0.4in}
  \centering
  \begin{overpic}[width=\linewidth]{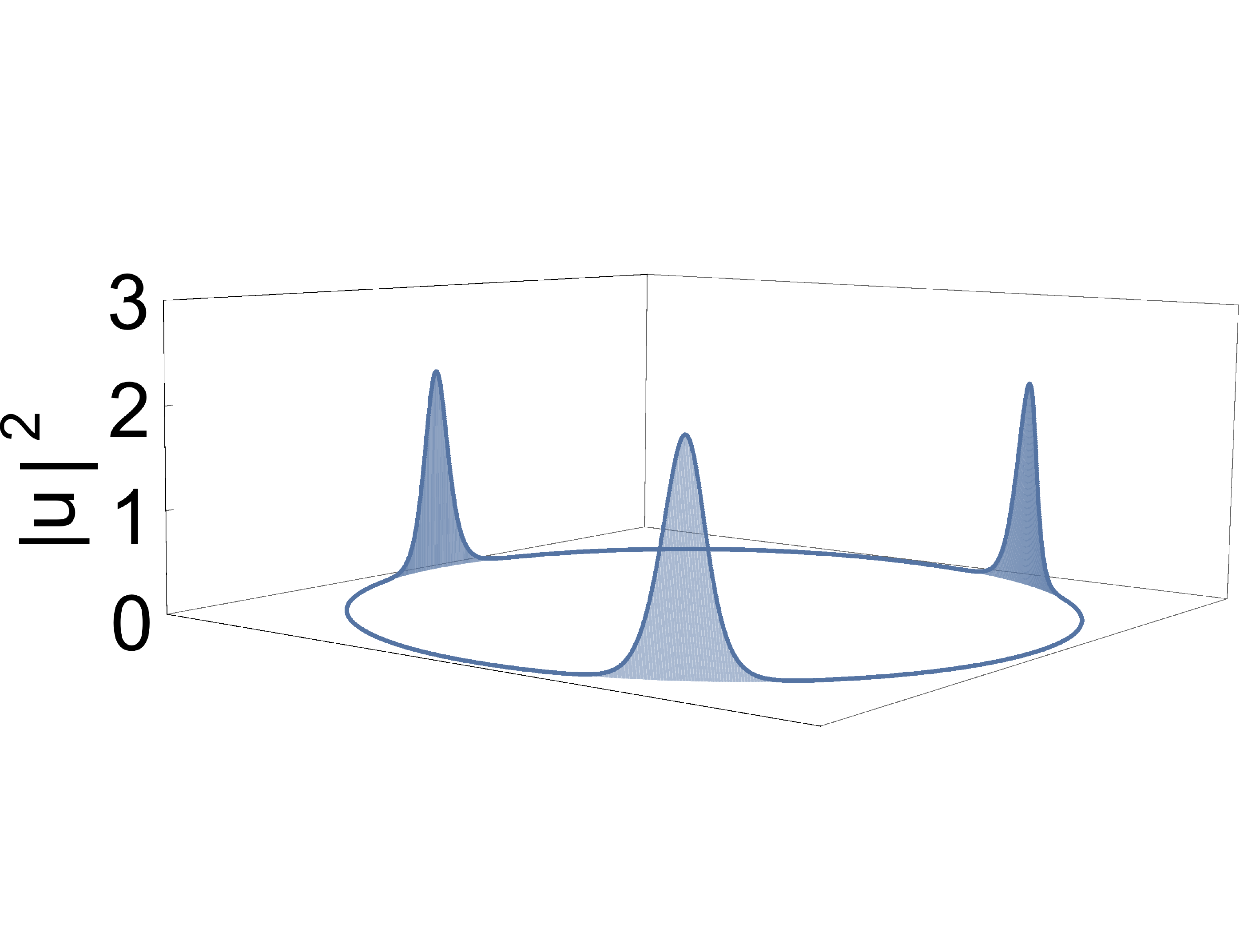}
  \put(17,76){(a)}
  \put(17,14){(c)}
  \end{overpic}
%  \caption{Number of solitons $n=3$}
  \label{fig:sfig3}
\end{subfigure}
\begin{subfigure}{.23\textwidth}
\vspace*{-0.2in}
  \centering
  \begin{overpic}[width=\linewidth]{4Dn_kLarge.pdf}
   \put(17,76){(b)}
   \put(17,14){(d)}
  \end{overpic}
% \caption{Number of solitons $n=4$}
  \label{fig:sfig4}
\end{subfigure}
\vspace*{-.40in}
\caption{The $\dn$ solution for $N=1, 2, 3$ and $4$. Here we plot $\dn$ solutions with the same modulus, $k^{2}=1-10^{-12}\approx 1$, as functions on a circle (to emphasize their periodicity). As the number of pulses $N$ increases, the width of the each pulse narrows and the height increases. These four solution branches co-exist for a fixed value of the detuning $\alpha$.  The stability of each solution branch depends upon the detuning parameter as shown in Fig.~\ref{fig:dn3}.}
\label{fig:dn2}
\end{figure}
Figure~\ref{fig:dn2} shows the $\dn$ solutions as the parameter $N$ is varied from one to four.  %{"Note that as $N$ is varied, the value of $k$ is also modified in order to satisfy the $2\pi$ periodic domain. Specifically, as $N$ increases, the value of $k\rightarrow 1$." Actually we didn't do this. With different $N$, we just varied $B$ to match with the period, but we can still choose $k$ to be anything. We are not keeping $B$ as a constant.} 

{Based on the observed behavior of these
  solutions of the LLE in numerical simulations,
  we consider solutions about the center frequency,
  $\xi=0$, and with a fixed phase term, which
  can be obtained by setting}
  \begin{equation}
    {\alpha = -{\beta}B^{2}(2-k^{2})/2.} \label{eq:alpha}
  \end{equation}
With these choices, the value of the detuning must be increased
in order to accommodate more pulses per round trip,
which is consistent with experimental findings.

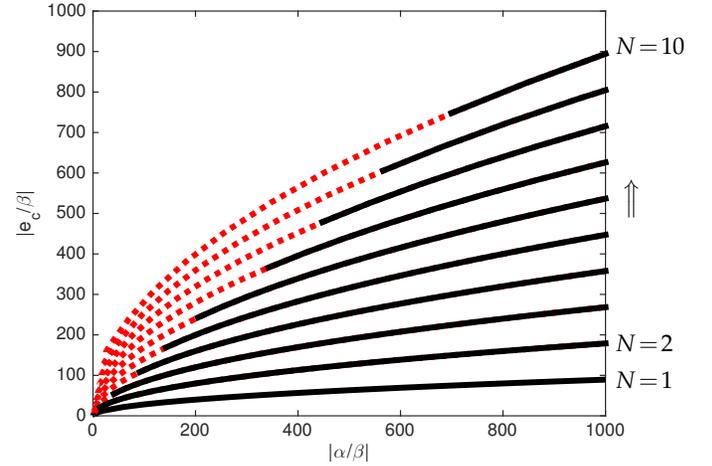
\begin{figure}
\begin{overpic}[width=\linewidth]{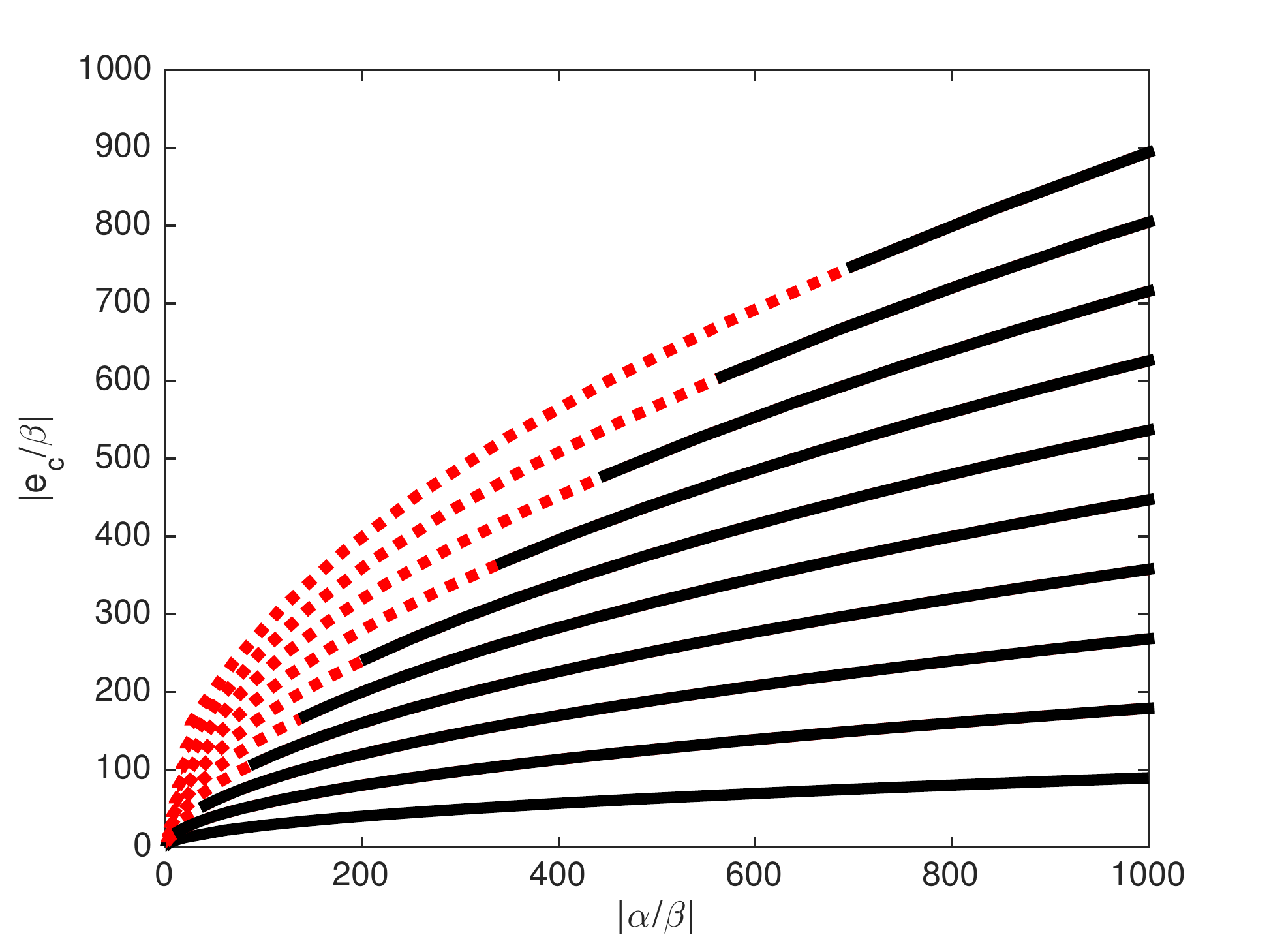}
\put(92,12.5){$N\!=\!1$}
\put(92,18){$N\!=\!2$}
\put(94,38){\rotatebox{90}{$\Longrightarrow$}}
\put(92,63){$N\!=\!10$}
\end{overpic}
\caption{The solution branches of the $\dn$ solution as a function of energy ($|e_c/\beta|$) versus detuning ($|\alpha/\beta|$).  Plotted are the solution branches from $N=1$ to $N=10$.  The instability of each branch can be computed from the linearized operator {Eq.~\ref{eq:Lmatrix}}. Specifically, if the real part of any eigenvalue crosses the threshold of $5\times 10^{-4}$, then the branch is considered unstable (dashed red lines) for that value of detuning.
For candidate branches that are potentially stable (black lines), further analysis (as shown in Sec.~5) is required to confirm the stability of the $\dn$ solution branches.  This figure matches recent experimental findings of \cite{kip2} and confirms that the specific number of {pulses} in the microresonator can be controlled by manipulation of the detuning.  }
\label{fig:dn3}
\end{figure}

Importantly, we can compute the cavity energy $e_c$ versus
detuning frequency $\alpha$ for the $\dn$ solutions by the
definition of the cavity energy as 
\begin{equation}
e_c=\int_{-\pi}^{\pi}|u_{0}|^{2}dx=-\beta B^{2}\int_{-\pi}^{\pi}\dn^{2}(By|k)dy \, .
\end{equation}
The energy of each solution branch can then be computed
for different $N$ values as shown in Fig.~\ref{fig:dn3}.
The stability of each branch will be discussed in what follows,
but the energy versus detuning shows the important trends to be
considered.
For $k \rightarrow 1$, the function $\dn(x|k) \rightarrow \sech(x)$
so that the energy integral can be approximated explicitly
\begin{equation}
{e_c \approx -\beta B\int_{-B\pi}^{B\pi}\sech^{2}zdz= -2\beta B \; .}
\end{equation}
{Given that $\alpha \approx -\beta B^{2}/2$, we can then} simplify the relationship between the detuning and cavity energy, i.e. $|e_c/\beta|\approx 2\sqrt{2}\sqrt{|\alpha/\beta|}$.
This value is for only a single {pulse}.  If there are $N$ {pulses},
we obtain
\begin{equation}
|e_c/\beta|=2\sqrt{2}N\sqrt{|\alpha/\beta|}.
\end{equation}
This gives a simple quantization of the energy as a function
of the number of pulses in the limit $k\rightarrow 1$.  We
will show in what follows that the $k\rightarrow 1$ limit is
where solutions to the LLE are stable, thus the energy quantization
formula is a good approximation for the LLE microresonator dynamics.
%In addition, the forcing $F$ in the LLE stabilizes the periodic solution with a shelf which is part of the experimentally and computationally observed solution.
{Note that in Fig.~\ref{fig:dn3}, since $|\alpha/\beta|=B^{2}/2$, we have $|\alpha/\beta| \rightarrow \infty$ when $k \rightarrow 1$ and $|\alpha/\beta| \rightarrow 0$ when $k \rightarrow 0$.}

\section{Stability Analysis of the LLE}

The stability of the Jacobi elliptic function solutions
to the LLE can be characterized using a linear 
stability analysis. Let $u_1 = e^{i\psi} w_1$.
Following the perturbation expansion
of \eqref{eq:perturb}, we find at leading order 
the Jacobi elliptic solutions and at $O(\epsilon)$ 
the linearized evolution
\begin{align}
\hat{F} &= i\frac{\partial w_1}{\partial t}-\alpha w_1
+2|u_{0}|^{2}w_1 -\frac{\beta}{2}\frac{\partial ^{2}w_1}{\partial x^{2}}
+|u_{0}|^{2}w_1^{*} \; ,
\end{align}
where $\hat{F}=i( e^{-i\psi} F  + e^{-i\psi}G(u_0,x,t) - \hat{u}_{0} 
- e^{-i\psi}u_{0\tau})$.

We can decompose the linearized evolution into real 
and imaginary components by letting 
$w_1=R+iI$ ($w_1^{*}=R-iI$) so that in matrix notation 
it takes the form
\begin{align}
\left[\!\!
\begin{array}{c}
R_{t}\\
I_{t}
\end{array}
\!\!\right] \!\!&=\!\!
\left[\!\!
\begin{array}{cc}
0 & 
\frac{\beta}{2}\partial^2_{x} -{{\hat u}_{0}}^{2}
+ \alpha\\
-\frac{\beta}{2}\partial^2_{x} +3{{\hat u}_{0}}^{2}
-\alpha & 
0
\end{array}
\!\!\right]
\!\!
\left[\!\!
\begin{array}{c}
R\\
I
\end{array}
\!\!\right]
\!+\!\left[ \!\!
\begin{array}{c}
\Im\hat{F}\\
\Re\hat{F}
\end{array}
\!\!\right]
\label{eq:spec} \; ,
\end{align}
where $\partial^2_x$ denotes the second order derivative.  
{The eigenvalue spectrum of the matrix in \eqref{eq:spec}
  yields the spectral stability of $\dn$ solutions, generally.
  Note that for ${\hat u}_{0}$ given by the $\dn$ solution,
  $\alpha=-{\beta}B^{2}(2-k^{2})/2$.}

%$\cn$ and $\sn$ solutions for which $\alpha=-{\beta}B^{2}(2-k^{2})/2$, $\alpha={\beta}B^{2}(1-2k^{2})/2$,
%and $\alpha={\beta}B^{2}(1+k^{2})/2$ respectively.

\begin{figure}[t]
\includegraphics[width=\linewidth]{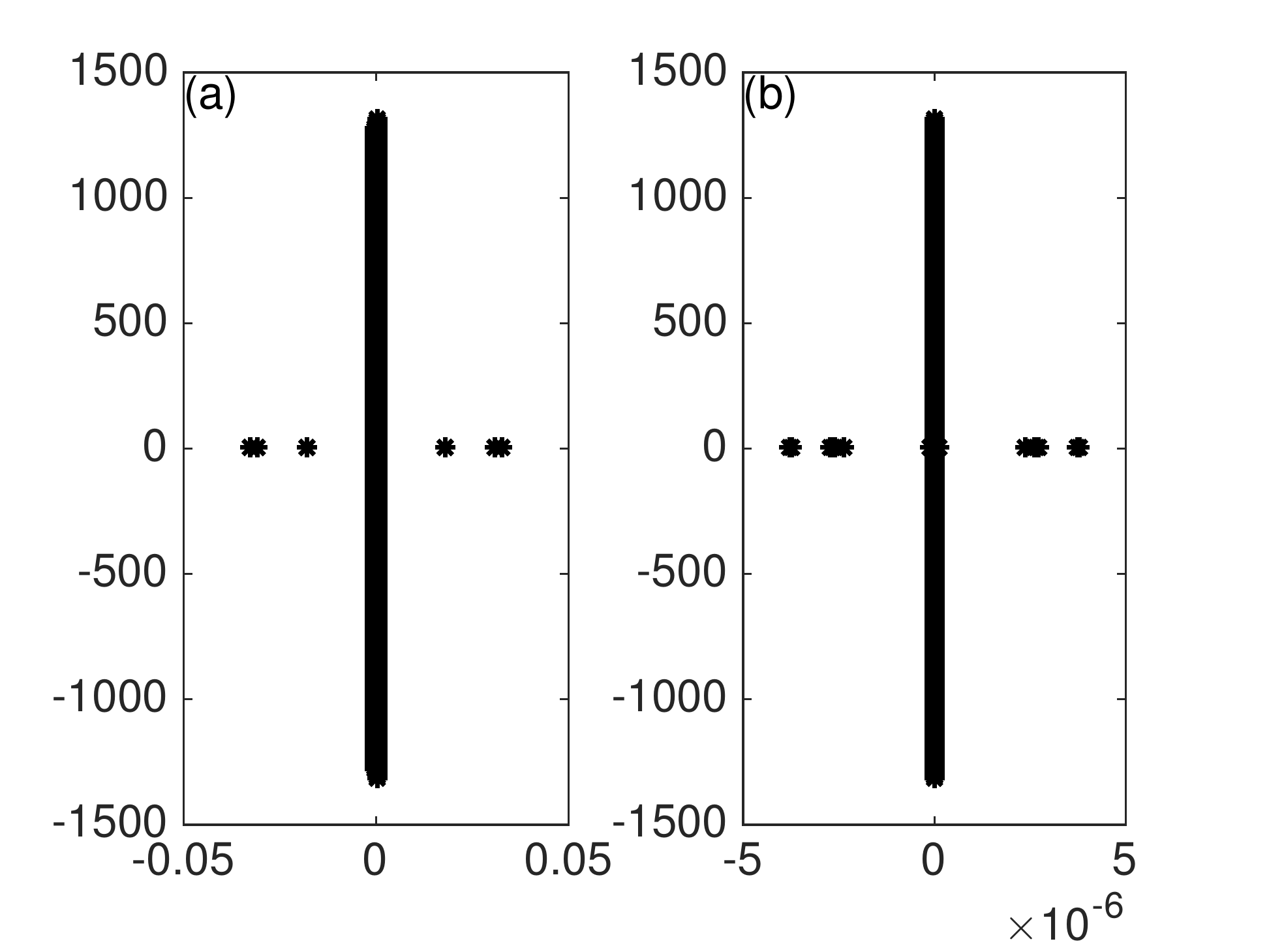}
\caption{Eigenvalue spectrum of the matrix ${\mathbf{L}}$ defined
  in \eqref{eq:spec} for the $\dn$ solution with (a) $k^2=0.9$ and (b) $k^{2}=1-10^{-12}\approx 1$. Although the eigenvalues shrink to the imaginary axis as $k\rightarrow 1$, the solutions are known to be unstable under generic perturbations.}
\label{fig:dn_spec}
\end{figure}

Figure~\ref{fig:dn_spec} shows the computed %--\ref{fig:sn_spec} 
spectrum of the linearized operator in \eqref{eq:spec}
for the $\dn$ solution with $N=4$. The operator was numerically 
evaluated using a spectrally accurate method with 1024
grid points (a fast Fourier transform was used to evaluate
the second derivatives) and a standard matrix eigenvalue
solver. The eigenvalues
corresponding to both $k^2=0.9$ and $k^2=1-10^{-12} \approx 1$
are evaluated. {Note for the case $N=4$, the
  fundamental period $T=\frac{2\pi}{4}$, thus $[-\pi, \pi]$ is
  a multiple of the fundamental period, so we expect instability
  \cite{stabilityDn}.} For $k^2=0.9$, {the $\dn$}
solution clearly has unstable eigenvalues, i.e. eigenvalues
with large positive real part.
As $k\rightarrow 1$, the real part of the eigenvalues {of $\dn$} shrink 
to the imaginary axis, suggesting that the $k\rightarrow 1$
solutions {will be better behaved, even if
  they are technically unstable \cite{stabilityDn}.}
Thus a critical part of the analysis is to determine if the 
addition of the LLE term $F$ stabilizes such microresonator 
solutions subject to slow-time modulation of the parameters.

\subsection{Discrete Spectrum of $\dn$ Solutions}

As with soliton perturbation theory, the generalized null space
of the adjoint of the linearized operator is critical for 
determining stability. Specifically, the Fredholm-alternative 
theorem stated in \eqref{eq:fredholm} requires that perturbations 
be orthogonal to the null space of the adjoint linear operator
(this removes so-called ``secular'' modes which have 
polynomial growth in time \cite{weinstein1985modulational}).  
For the $\dn$ solution, the linear operator reduces to
\begin{equation}
  {\bf L}=-\frac{\beta}{2}B^2 \left[  \begin{array}{cc}  0 & L_- \\ -L_+ & 0  \end{array}  \right],
\label{eq:Lmatrix}
\end{equation}
with the self-adjoint operators
\begin{subeqnarray}
  &&  L_- = -\frac{d^2}{dz^2} -2 \dn^2 z - k^{2} + 2 \; ,\\
  &&  L_+ = -\frac{d^2}{dz^2} - 6 \dn^2 z - k^{2} + 2 \; ,
\end{subeqnarray}
where a change of variables to $z=B(x-x_0)$ has been made
and the dependence of $\dn$ on $k$ has been dropped for
notational convenience.

In the following, we 
will denote the generalized nullspace of a linear operator 
$L$ by $\ker_g(L)$, i.e.

\begin{equation}
  \ker_g(L) = \bigcup_{m=1}^\infty \ker(L^m) \; .
\end{equation}
We also require the space {$H^m_\per[a,b)$},
  which denotes 
  a periodic Sobolev space on {$[a,b)$}.
    This space
may be characterized by the Fourier coefficients of a 
given function. Let $f$ be a function {defined
on $[a,b) = [-\pi,\pi)$}
and let $c_j$ defined by 

\begin{equation}
c_j = \frac{1}{2\pi} \int_{-\pi}^\pi f(x) e^{-i j x} \, dx \; .
\end{equation}
Then the {$H^m_\per[-\pi,\pi)$} norm is defined by

\begin{equation}
 { \| f \|^2_{H^m_\per} = \sum_j |c_j|^2 (1+|j|^2+\cdots |j|^{2m}) \; .}
\end{equation}
Note that if {$\| f\|_{H^m_\per} < \infty$}, then 
the Fourier coefficients of the {$(m-1)st$ derivative
of} $f$ are absolutely
summable so that {$f^{(m-1)}$} is continuous as a periodic 
function on $[-\pi,\pi)$. The space {$H^m_\per$} can
  be defined for other intervals by appropriate
  scaling.

Let $w_1 = R + i I$ as above.
As in \cite{weinstein1985modulational}, we define 
the space 
\begin{equation}
  \mathcal{M} = H^1_\per \times H^1_\per
      \bigcap \left (\ker_g( {\bf L}^\dag ) \right )^\perp \; ,
\end{equation}
which is where we will constrain the evolution
of $(R,I)^\intercal$. {Note that the domain for $z$
is $[-NK(k),NK(k))$}. We also define the
periodic functions
$\phi(z)$ and $\varphi(z)$ to be 
\begin{subeqnarray}
&& \hspace*{-.5in} \phi(z)=(K(k)E(z,k)- E(k)z) \dn z-k^{2}K(k)\sn z\cn z,\\
  && \hspace*{-.5in} \varphi(z)=k^{2} \cn z \sn z (K(k)E(z,k)-E(k)z) \nonumber\\
    && \hspace*{-.5in} \qquad \qquad +(E(k)-K(k))
    \dn z+k^{2}K(k)\cn^{2} z\dn z \; ,
\end{subeqnarray}
where {$E(z,k)$}$=\int_{0}^{z}\dn^{2} ydy$ is the incomplete
elliptic integral of the second kind, $E(k) = E(K(k),k)$
is the complete elliptic integral of the second kind,
and $K(k)$ is as above. For the sake of compactness,
we will often drop the dependence
of $E(k)$ and $K(k)$ on the modulus $k$ in the following.
{Note that $E(z,k)$ is odd,
  $\phi(z)$ is odd, and $\varphi(z)$ is even --- the
  parity of functions simplifies much of the
  following analysis.}
A set of eigenfunctions that span $\ker_g({\bf L}^\dag)$
can then be computed from the following observations

\begin{subeqnarray}
&& \hspace*{-.5in} L_{-}[\dn z]=0,\\
&& \hspace*{-.5in} L_{+}[\sn z\cn z]=0,\\
%&& \hspace*{-.5in} L_{+}  L_{-} \left [\frac{1}{2k^{2}}z\dn z \right ]
&& \hspace*{-.5in}{L_{+}  L_{-} \left [\phi(z) \right ]
= L_{+} [-2k^{2}E\sn z\cn z] = 0,}\\
%&& \hspace*{-.5in} L_{-}  L_{+} \left [\frac{1}{2(2-k^{2})}(k^{2}z\sn z\cn z
%-\dn z) \right]= L_{-} [\dn z]=0.
&& \hspace*{-.5in} {L_{-}  L_{+} \left [\varphi(z) \right]=L_{-} [2((k^{2} - 2)E-2(k^{2}-1)K)\dn z]=0,} \nonumber \\
%&& \hspace*{-.5in} {L_{-} [2((k^{2} - 2)E-2(k^{2}-1)K)\dn z]=0,}
\end{subeqnarray}

These results are used to derive some important
properties of the operators ${\bf L}^\dag$, $L_+$, and
$L_-$, which are summarized in propositions
\ref{prop:lplmnull} and \ref{prop:geneig}. Proofs are
included in the appendix.
\begin{proposition} \label{prop:lplmnull}
  {Assume $N\in \mathbb{N}$ and} $0<k<1$.
  The operator $L_-$ is non-negative and self-adjoint, with
  $\ker(L_-) = \vecspan \{ \dn z \}$.
  The operator $L_+$ is self-adjoint, with
  $\ker(L_+) = \vecspan \{ \sn z \cn z \}$.
\end{proposition}
\begin{proposition} \label{prop:geneig}
  {Assume $N\in \mathbb{N}$ and}
  $0<k<1$ and let $(f,g)^\intercal \in H^1_\per \times H^1_\per$.
  If the following orthogonality relations hold
  \begin{subeqnarray}
    \langle f, \dn z \rangle &= 0 \; , \\
    \langle f,  {\phi(z)} \rangle &= 0 \; , \\   
    \langle g, \sn z \cn z \rangle &= 0 \; , \\
    \langle g, {\varphi(z)} \rangle &= 0 \; ,
  \end{subeqnarray}
  then $(f,g)^\intercal \in \mathcal{M}$.
\end{proposition}

\subsection{Bounding the evolution of $w_1$ {($N=1$)}}

Following the analysis of Weinstein~\cite{weinstein1985modulational}, the evolution of
the term $w_1 = R + iI$ is bounded by considering the
function
\begin{equation}
Q(f,g) =-\frac{\beta}{2}B^{2}[\langle L_{+}f,f \rangle 
  + \langle L_{-}g,g \rangle ] \; ,
\end{equation}
which is a conserved quantity along the solution
trajectory for $w_1$, i.e. $dQ(R,I)/dt=0$. For
$(R,I)^\intercal \in \mathcal{M}$, we have the following bound.

\begin{proposition} \label{prop:h1control}
  {Assume $N=1$}. Let $w = R +iI \in \mathcal{M}$. Then
  there exist constants $C_1$ and $C_2$ such that
  \begin{eqnarray}
    C_1 \left ( \|R\|^2_{H^1_\per} + \|I\|^2_{H^1_\per} \right) &\leq&
    Q(R,I) \; ,\\
    C_2 \left ( \|R\|^2_{H^1_\per} + \|I\|^2_{H^1_\per} \right) &\geq&
    Q(R,I) \; .
  \end{eqnarray}
\end{proposition}
This proposition is the primary result needed in
our analysis: if the slow evolution of the parameters
$B$, $\xi$, $x_0$, and $\sigma$ is such that
$(R(t),I(t))^\intercal \in \mathcal{M}$, then for any $T_0$ we have
{$\sup_{0 \leq t \leq T_0/\epsilon}$} $\| \epsilon w_1(t) \|_{H^1_\per}
\to 0$ as $\epsilon \to 0$. See
\cite{weinstein1985modulational} for details.  A proof of
Proposition~\ref{prop:h1control} based on a variational
formulation can be found in \cite{stabilityDn}. A more
classical proof modeled after \cite{weinstein1985modulational}
is provided in the appendix.

\subsection{Modulation equations for the $\dn$ solution
{($N=1$)}}

For the $\dn$ solution,
\begin{align}
  \partial_\tau \left (\hat{u}_0 \right) &= \sqrt{-\beta} B_{\tau}\dn z
  + A \frac{dk}{dB} B_\tau \dn_{k} z \nonumber\\
  &\qquad+ A [B_{\tau}(x-x_{0})-Bx_{0\tau}] \dn_{z} z\; .
\end{align}
We consider solutions with $k \rightarrow 1$, so
that $dk/dB \approx 0$.
Using this approximation, the above reduces to

\begin{align}
  \partial_\tau \left (\hat{u}_0 \right) &= \sqrt{-\beta}B_{\tau} \dn z- Ak^{2}
  \sn z \cn z \left (\frac{B_{\tau}}{B}z-Bx_{0\tau} \right ) \; ,
\end{align}
so that 

\begin{align}
\partial_\tau \left(  u_0 \right ) &= 
e^{i\psi} \sqrt{-\beta}B_{\tau} \dn z- e^{i\psi}Ak^{2}
\sn z \cn z \left (\frac{B_{\tau}}{B}z-Bx_{0\tau} \right ) 
\nonumber\\
&\qquad +i \left (\xi_\tau \frac{z}{B} - \xi x_{0\tau} - \sigma_{0\tau}
\right ) e^{i\psi} A \dn z
\end{align}
This gives the following expression for the forcing term

\begin{align}
    \Im\hat{F} &= F \cos \psi + \Re (e^{-i\psi}G)
    -\sqrt{-\beta}B_{\tau} \dn z - A \dn z \nonumber \\
    & \qquad +Ak^{2} \left (\frac{B_{\tau}}{B}z
    -Bx_{0\tau} \right ) \sn z \cn z,\\
    \Re\hat{F} &= F\sin \psi -\Im (e^{-i\psi}G)
    + \sqrt{-\beta}\xi_{\tau}z \dn z \nonumber \\
    & \qquad -(\xi x_{0\tau}+\sigma_{0\tau})A \dn z \; .
\end{align}

To constrain the forcing term $(\Im\hat{F}, \Re\hat{F})^\intercal$
to be in $\mathcal{M}$, Proposition~\ref{prop:geneig} implies
the following constraints

\begin{subeqnarray}
&& \hspace*{-.5in} \langle\Im\hat{F},\dn z\rangle = 0,\\
%&& \hspace*{-.5in} \langle\Im\hat{F},z\dn z\rangle = 0,\\
&& \hspace*{-.5in} \langle\Im\hat{F},{\phi(z)} \rangle = 0,\\
&& \hspace*{-.5in} \langle\Re\hat{F},\sn z\cn z\rangle = 0,\\  
%&& \hspace*{-.5in} \langle\Re\hat{F},k^{2}z\sn z\cn z-\dn z\rangle = 0.
&& \hspace*{-.5in} \langle\Re\hat{F},{\varphi(z)} \rangle = 0.
\end{subeqnarray}
These constraints require the slow evolution of the parameters
to satisfy the following system of differential equations
\begin{subeqnarray}
&&\hspace*{-.5in}\frac{dB}{d\tau}=\frac{\langle\Re(e^{-i\psi}G)+F\cos\psi,\dn z\rangle-\sqrt{|\beta|}B\langle\dn z, \dn z \rangle}{\sqrt{|\beta|}(\langle \dn z,\dn z\rangle-k^{2}\langle z\sn z \cn z, \dn z\rangle)} \; , \slabel{eq:pert2a} \\
&&\hspace*{-.5in}{\frac{dx_{0}}{d\tau}=\frac{\langle F\cos\psi +\Re(e^{-i\psi}G), \phi(z) \rangle}{\sqrt{|\beta|}B^{2}k^{2}\langle \sn z\cn z, \phi(z) \rangle}} \; ,\\
  &&\hspace*{-.5in}\frac{d\xi}{d\tau}=-\frac{\langle F\sin\psi -\Im(e^{-i\psi}G), \sn z\cn z\rangle}{\sqrt{|\beta|}\langle z\dn z, \sn z\cn z\rangle}  \; , \\
  &&\hspace*{-.5in}{\frac{d\sigma_{0}}{d\tau}+\xi\frac{dx_{0}}{d\tau}=\frac{\langle F\sin\psi -\Im(e^{-i\psi}G), \varphi(z) \rangle}{\sqrt{|\beta|}B\langle \dn z, \varphi(z) \rangle}}\, ,
%&& \qquad\qquad-\frac{\Im(e^{-i\psi}G), \varphi(z) \rangle}{\sqrt{|\beta|}B\langle \dn z, \varphi(z) \rangle} \, ,
\label{eq:pert2}
\end{subeqnarray}
where $\psi= {\xi}z/B+\sigma-\sigma_{0}$.
We can further simplify the equations above by applying
trigonometric identities, we have
\begin{subequations}
\begin{align}
&\frac{dB}{d\tau}=\frac{\langle\Re(e^{-i\psi}G),\dn z\rangle+F\cos(\sigma-\sigma_{0})p_{1}(\xi)}{\sqrt{|\beta|}(q_1(k)-k^2q_2(k))} \nonumber\\
& \qquad\qquad -\frac{\sqrt{|\beta|}B\langle\dn z, \dn z \rangle}{\sqrt{|\beta|}(q_1(k)-k^2q_2(k)} \; , \slabel{eq:pert2a} \\
&{\frac{dx_{0}}{d\tau}=\frac{\langle \Re(e^{-i\psi}G), \phi(z) \rangle -F\sin(\sigma-\sigma_{0})p_{2}(\xi)}{\sqrt{|\beta|}B^{2}k^2q_3(k)}} \; ,\\
&\frac{d\xi}{d\tau}=\frac{\langle \Im(e^{-i\psi}G), \sn z\cn z\rangle-F\cos(\sigma-\sigma_{0})p_{3}(\xi)}{\sqrt{|\beta|}q_2(k)}  \; , \\
&{\frac{d\sigma_{0}}{d\tau}+\xi\frac{dx_{0}}{d\tau} =\frac{F\sin(\sigma-\sigma_{0})p_{4}(\xi)-\langle \Im(e^{-i\psi}G), \varphi(z)\rangle}{\sqrt{|\beta|}Bq_4(k)}} \, , \nonumber \\
%& \qquad\qquad\qquad - \frac{\langle \Im(e^{-i\psi}G), k^{2}z\sn z\cn z-\dn z\rangle}{\sqrt{|\beta|}B(q_2(k)-q_1(k))} \, ,
\label{eq:pert2}
\end{align}
\end{subequations}
where

\begin{subequations} \label{eq:xiterms}
\begin{align}
  p_1(\xi) &= \langle \cos(\xi z/B), \dn z \rangle \; ,\\
  p_2(\xi) &= {\langle \sin(\xi z/B), \phi(z) \rangle} \; ,\\
  p_3(\xi) &= \langle \sin(\xi z/B), \sn z \cn z \rangle \; ,\\
  p_4(\xi) &= {\langle \cos(\xi z/B), \varphi(z) \rangle} \; ,
\end{align}
\end{subequations}
and 
\begin{subequations}
\begin{align}
  q_1(k) &= \langle \dn z,\dn z\rangle \\
  q_2(k) &= \langle z \dn z,\sn z \cn z\rangle \\
  q_3(k) &= {\langle \sn z\cn z, \phi(z)\rangle} \\
  q_4(k) &= {\langle \dn z, \varphi(z)\rangle} \; .
\end{align}
\end{subequations}

In addition to these constraints, $\xi$ should be an
integer so that $u_0$ remains in $H^1_\per$. Therefore
the analysis is only rigorous when applied to
perturbations for which $d\xi/d\tau = 0$, but we
have found that the analysis provides insight in other
cases.

Consider the stability of this system of
differential equations
around the center frequency, i.e. $\xi=0$, so that
$\psi= \sigma-\sigma_{0}$ and $p_{2}(\xi)=p_{3}(\xi)=0$.
For $k \approx 1$, we can approximate many of the 
inner products in the above expressions using the 
limiting forms of the Jacobi elliptic functions. 
We obtain the approximations $p_1(0) \approx \pi$, 
$p_4(0) \approx 0$, $q_1(k) \approx 2$, 
$q_2(k) \approx 1$, {$q_3(k) \approx -1$,
  and $q_4(k) \approx 1$.}
With these approximations, the 
evolution equations simplify to

\begin{subeqnarray}
&&\hspace*{-.5in}\frac{dB}{d\tau}=\frac{1}{\sqrt{|\beta|}}(\langle\Re(e^{-i\psi}G),\dn z\rangle+F\pi \cos(\sigma-\sigma_{0}))-2B \; , \slabel{eq:pert2a} \\
&&\hspace*{-.5in}\frac{dx_{0}}{d\tau}=-\frac{\langle\Re(e^{-i\psi}G),{\phi(z) \rangle}}{{\sqrt{|\beta|}B^{2}}} \; ,\\
&&\hspace*{-.5in}\frac{d\xi}{d\tau}=\frac{1}{\sqrt{|\beta|}}\langle\Im(e^{-i\psi}G),\sn z\cn z\rangle \; , \\
&&\hspace*{-.5in}\frac{d\sigma_{0}}{d\tau}+\xi\frac{dx_{0}}{d\tau}={-\frac{\langle\Im(e^{-i\psi}G),\varphi(z)\rangle}{\sqrt{|\beta|}B}} \, .
\label{eq:pert2}
\end{subeqnarray}
These slow evolution equations approximate the
effect of a perturbation $G$ on the microresonator
comb.

\subsection{Stability of $\dn$ solutions of the LLE
  {($N=1$)}}

When the perturbation $G=0$, the parameter evolution constraints
\eqref{eq:pert2} yield the following set of equations
\begin{subeqnarray}
&&  \frac{dB}{d\tau}=\frac{F\pi \cos(\sigma-\sigma_{0})}{\sqrt{|\beta|}}-2B  \; \slabel{eq:dbdtau},\\
&& \frac{dx_{0}}{d\tau}=0 \; ,\\
&& \frac{d\xi}{d\tau}=0 \; ,\\
&&\frac{d\sigma_{0}}{d\tau}+\xi\frac{dx_{0}}{d\tau}=0 \; ,
\end{subeqnarray}
which gives the solution with $B=F\pi \cos(\sigma-\sigma_{0})/(2 \sqrt{|\beta|})$ as a steady-state attractor to the dynamics.  Specifically, values of $B$ larger and smaller than this exponentially decay back to the steady-state value.   In addition, the fast time scale dynamics give
\begin{subeqnarray}
&& \frac{dx_{0}}{dt}=-\beta \xi, \\
&& \frac{d\sigma}{dt}=-\alpha-\frac{\beta}{2}B^{2}(2-k^{2})-\frac{\beta}{2}\xi^{2}.
\end{subeqnarray}
Integrating the constant $\sigma_0$ into the second equation and setting $\xi=0$, i.e. we are working around the center frequency, then
\begin{equation}
\frac{d(\sigma-\sigma_{0})}{dt}=-\alpha-\frac{\beta}{2}B^{2}(2-k^{2}).
\end{equation}
Since the first solvability condition gives the steady-state $B={F}\pi \cos(\sigma-\sigma_{0})/({2\sqrt{-\beta}})$, then
\begin{equation}
\frac{d(\sigma-\sigma_{0})}{dt}={-\alpha+\frac{F^{2}\pi^{2}(2-k^{2})}{8}\cos^{2}(\sigma-\sigma_{0})} \; ,
\end{equation}
%
%{$\frac{F^{2}\pi^{2}}{8}\cos^{2}(\sigma-\sigma_{0})$ This is not true! Should be $-\alpha+\frac{F^{2}\pi^{2}(2-k^{2})}{8}\cos^{2}(\sigma-\sigma_{0})$!}
%

and 
\begin{equation} \label{eq:phase_predict}
{\cos(\sigma-\sigma_{0}) =  \frac{2B\sqrt{-\beta}}{F\pi}} \; ,
\end{equation}
which gives the time-independent phase of the microresonator comb.  Specifically, the real part of the solution is $u_{0}=A\dn(B(x-x_{0}),k)\cos(\sigma-\sigma_{0})$ and the imaginary part is $u_{0}=A\dn(B(x-x_{0}),k)\sin(\sigma-\sigma_{0})$.   

These asymptotic results show that $B\neq 0$ provided $F>0$.
Moreover, the stable microresonator solution has a fixed
phase relation which does not evolve in time.  Simulations show
that these two predictions are accurate representations of the
dynamics.  More than that, the prediction here shows them to
be attractors for general initial conditions, which is again
borne out by simulation.

\begin{figure}[t]
\includegraphics[width=\linewidth]{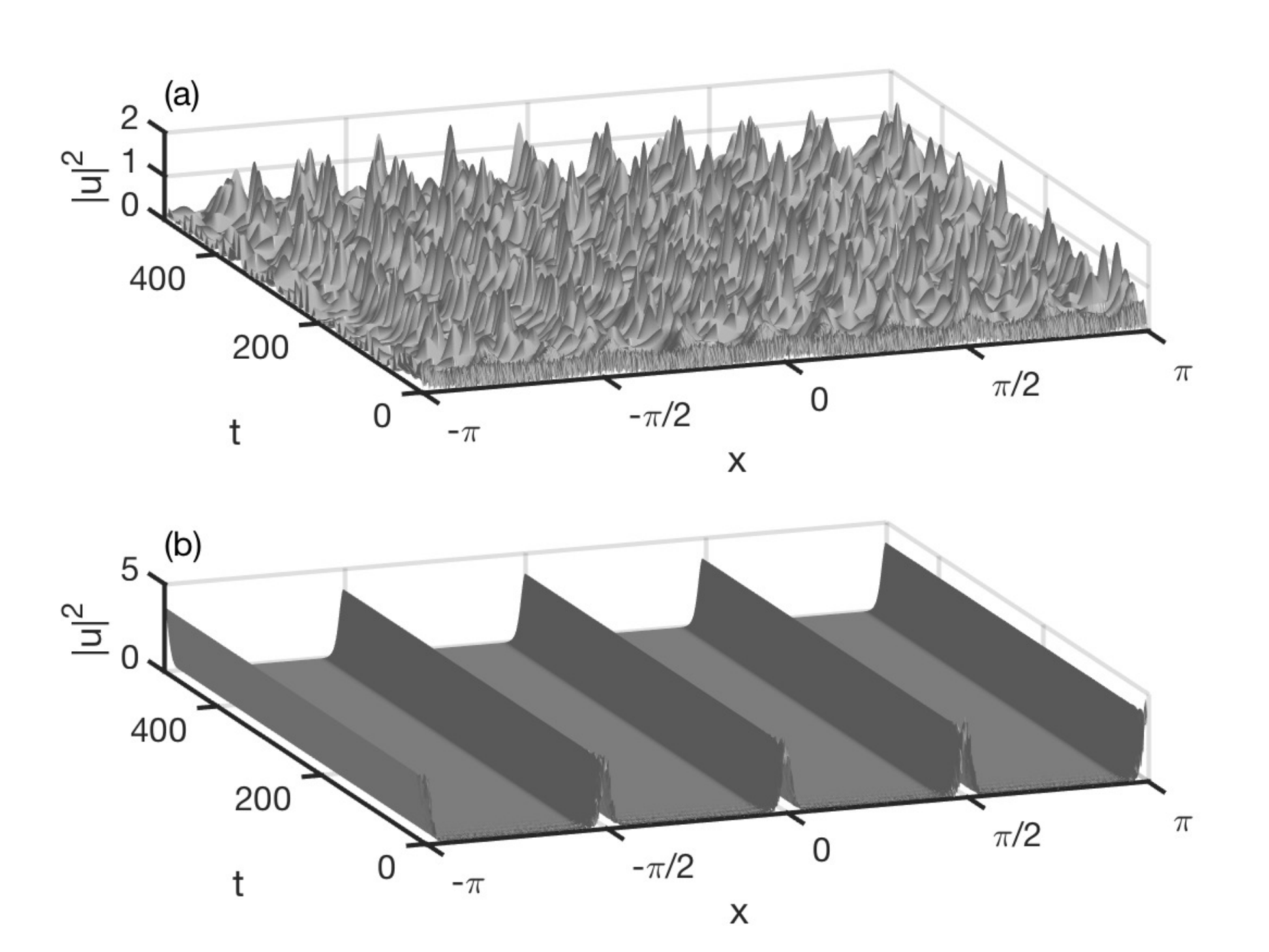}
\caption{Numerical simulation {of the ($N=4$) $\dn$ solution of \eqref{eq:lle} with $\epsilon=0.1$, $G=0$, and the detuning $\alpha$ set to (a) $\alpha = 0.0593$ and (b) $\alpha = 1.8732$ so that the appropriate parameter for the initial waveform is (a) $k^2=0.9$ and (b) $k^{2}=1-10^{-12}\approx 1$, according to \eqref{eq:alpha}}. The {initial waveform was corrupted with white noise} to induce instability in the evolution. The $k^2=0.9$ solution is shown to be unstable whereas the $k^2=1-10^{-12}$ solution is stable.  This is consistent with our linear stability analysis and Fig.~\ref{fig:dn3}.}
\label{fig:dn_evolve1}
\end{figure}

\begin{figure}[t]
\includegraphics[width=\linewidth]{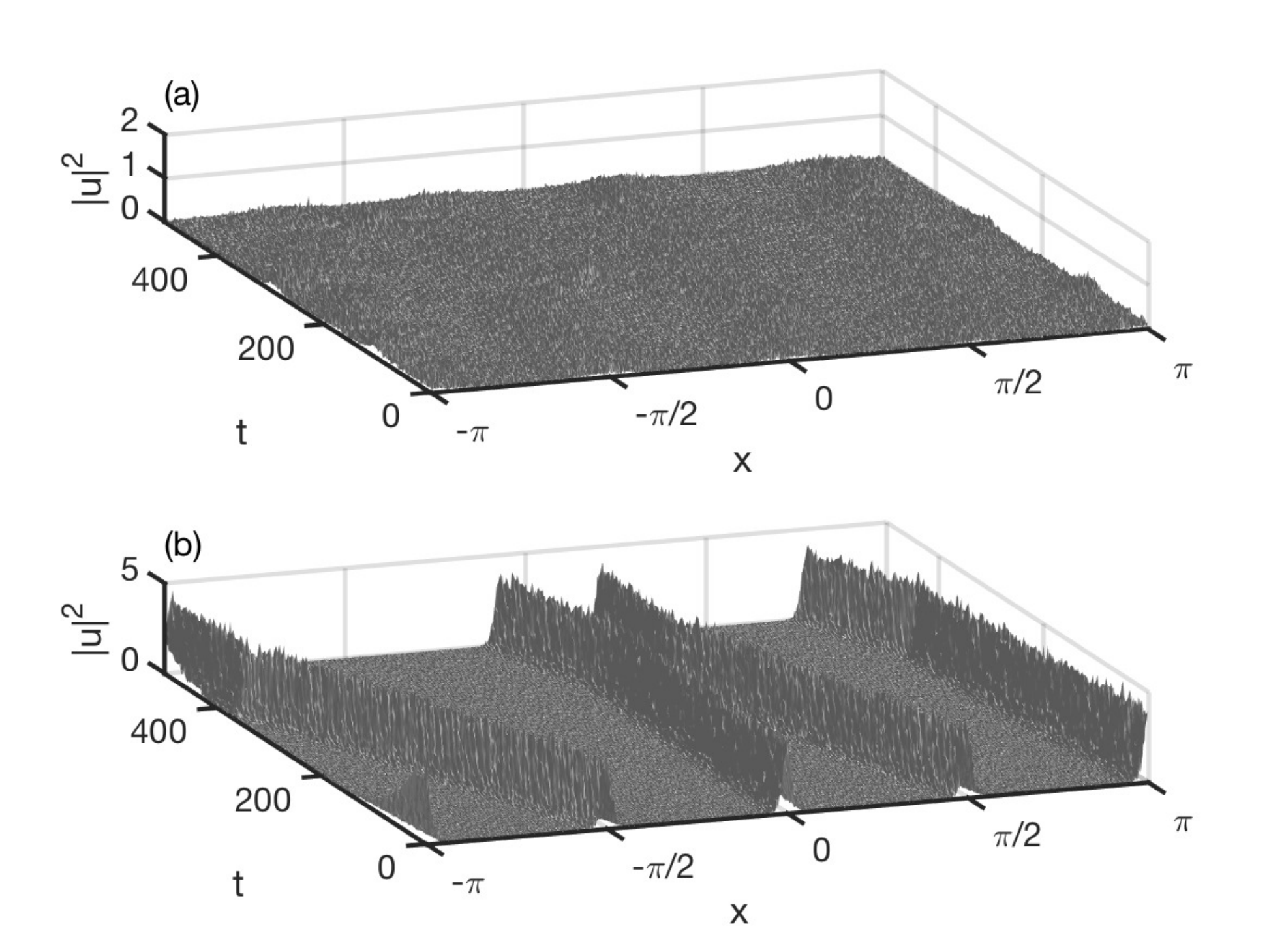}
\caption{{Numerical simulation of the ($N=4$) $\dn$ solution of \eqref{eq:lle} with $\epsilon=0$, $G=0$, and the detuning $\alpha$ set to (a) $\alpha = 0.0593$ and (b) $\alpha = 1.8732$ so that the appropriate parameter for the initial waveform is (a) $k^2=0.9$ and (b) $k^{2}=1-10^{-12}\approx 1$, according to \eqref{eq:alpha}}. {The initial waveform was corrupted with the same white noise as Fig.~\ref{fig:dn_evolve1} to induce instability in the evolution. Both the $k^2=0.9$ solution and the $k^2=1-10^{-12}$ solution are unstable.} {Comparing with Fig.~\ref{fig:dn_evolve1}, we observe that the $k^{2}=1-10^{-12}$ solution is stabilized by the LLE type perturbation.}}
\label{fig:dn_evolve0}
\end{figure}

{\subsection{The case $N>1$ in practice}

  Proposition~\ref{prop:lplmnull} states that 
  the nullspaces of $L_+$ and $L_-$ are each spanned
  by one function, for any positive number of pulses $N$.
  However, in simulations of finite precision, this
  mathematical truth is not observed for $k\approx 1$.
  Indeed, the discretizations of these operators
  are observed to have $N$ eigenfunctions corresponding to
  a zero (to numerical precision) eigenvalue when
  $k\approx 1$. Intuitively, this results from the
  fact that the $\dn$ function is nearly zero between
  pulses for $k\approx 1$ so that the pulses are essentially
  decoupled. Indeed, the set of $N$ shifted copies of the
  eigenfunctions for $N=1$, i.e. individual pulses in
  each of $[-K,K),[K,3K),\ldots,[(2N-3)K,(2N-1)K)$, is seen
      to give a basis for these nullspaces, again to
      numerical precision.

      Counterintuitively, it is this failure of
      Proposition~\ref{prop:lplmnull} in practice which
      explains the predictive power of the modulation
      equations of the previous sections --- which hold
      mathematically only when $N=1$ --- for numerical
      simulations with $N>1$. In particular, for the
      LLE type perturbation alone ($G=0$), we observe
      that the stabilizing effect on the
      amplitude of the comb as predicted by \eqref{eq:dbdtau}
      and the generation of a time independent phase as
      predicted by \eqref{eq:phase_predict}
      for the $N=1$ case also hold for $N>1$ when
      $k\approx 1$. See
      Figures~\ref{fig:dn_evolve1} and \ref{fig:dn_evolve0}
      for a comparison of the stability of a $\dn$
      initial condition with and without the LLE terms,
      which we discuss in more detail in the next section.
      For nonzero $G$ perturbations, as in the
      Raman effect and spontaneous emission noise
      examples below, the more qualitative $N=1$ predictions
      are also observed numerically when $N>1$. The fact
      that the behaviors of the
      pulses have decoupled is only apparent for the
      spontaneous emission noise example, as the perturbations
      acting on each pulse are identical in the other
      examples.
  
}

\begin{figure}[t]
\includegraphics[width=\linewidth]{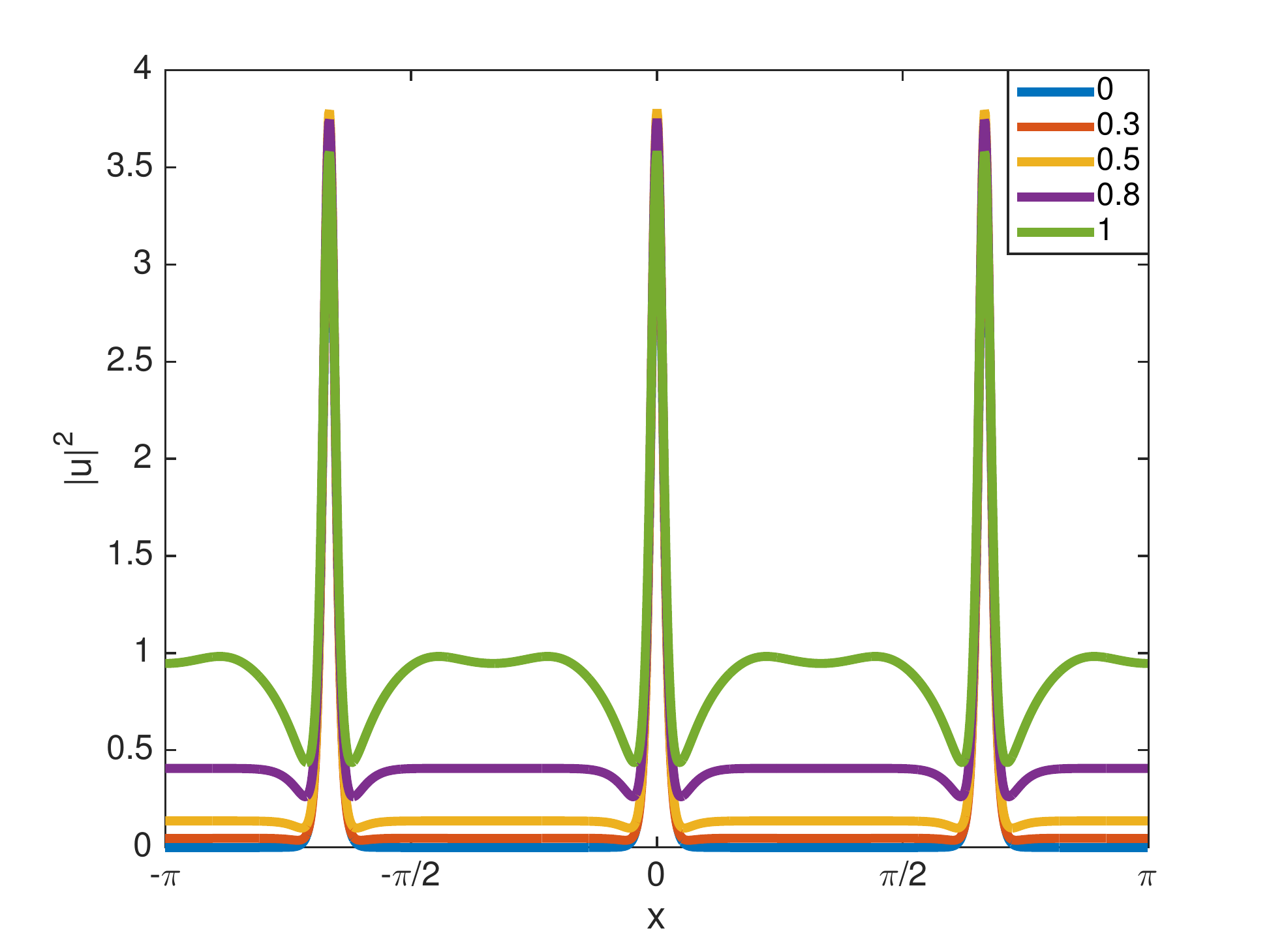}
\caption{Stable numerical solutions of {\eqref{eq:lle} with $G=0$ and $\alpha=1.7793$ for various values of $\epsilon$.} {The initial waveform was set as a ($N=3$) $\dn$ solution with $k^2=1-10^{-16}$.} As $\epsilon$ is increased from $\epsilon\ll 1$, the solutions deform from the asymptotic $\dn$ form to a localized structure that sits atop a shelf.  Importantly, like the $\dn$ solution, the resulting evolution produces solutions which have no nodal separation between neighboring pulses. }
\label{fig:dn_evolve2}
\end{figure}

\begin{figure}[t]
\begin{subfigure}{.25\textwidth}
%  \centering
\hspace*{-.2in}
  \begin{overpic}[width=\linewidth]{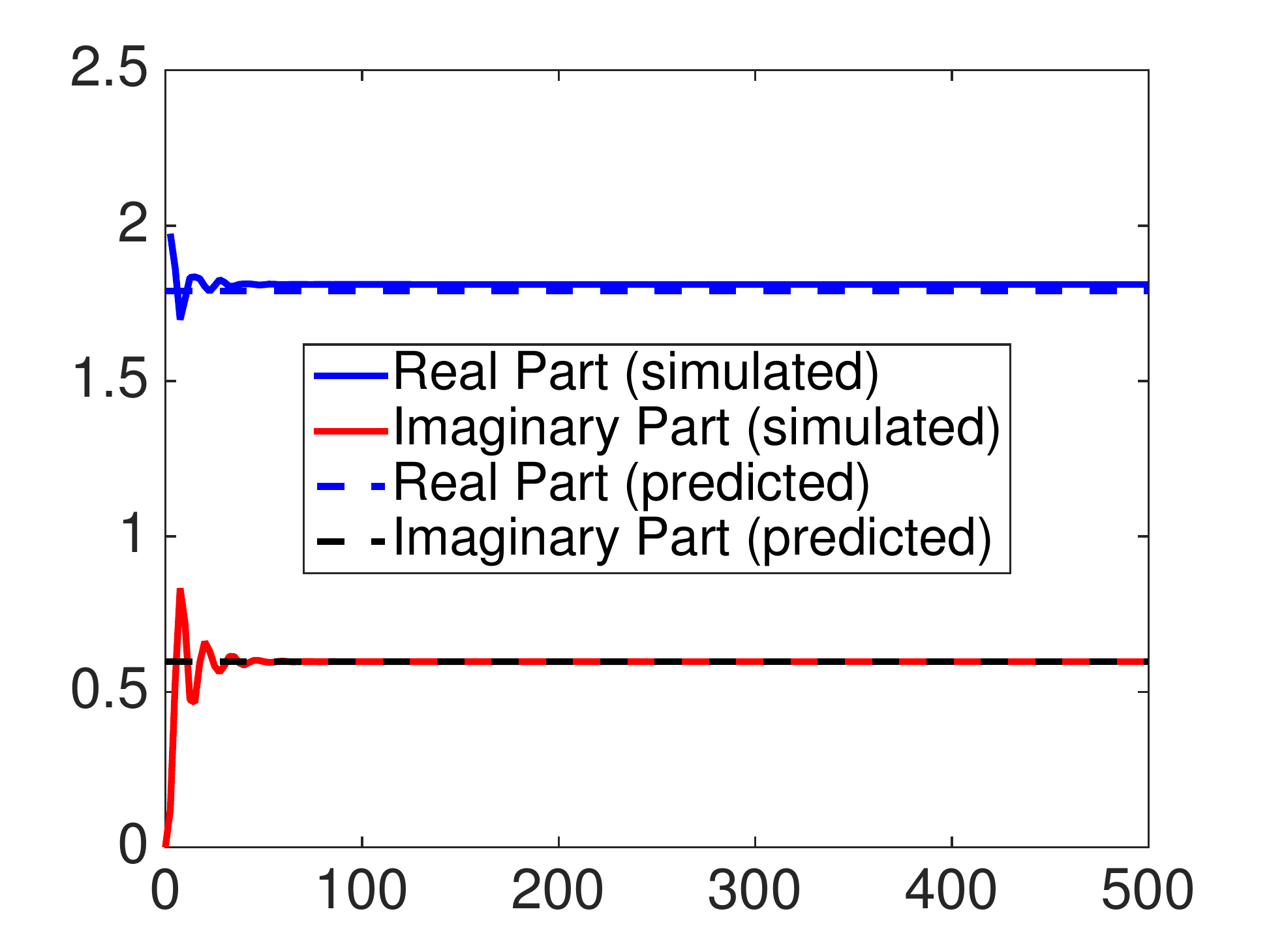}
  %\caption{$\epsilon=0.1$}
\put(19,13){(a) $\epsilon=0.1$}
\end{overpic}
%  \label{fig:sfig1}
\end{subfigure}%
\begin{subfigure}{.25\textwidth}
 % \centering
 \hspace*{-.3in}
  \begin{overpic}[width=\linewidth]{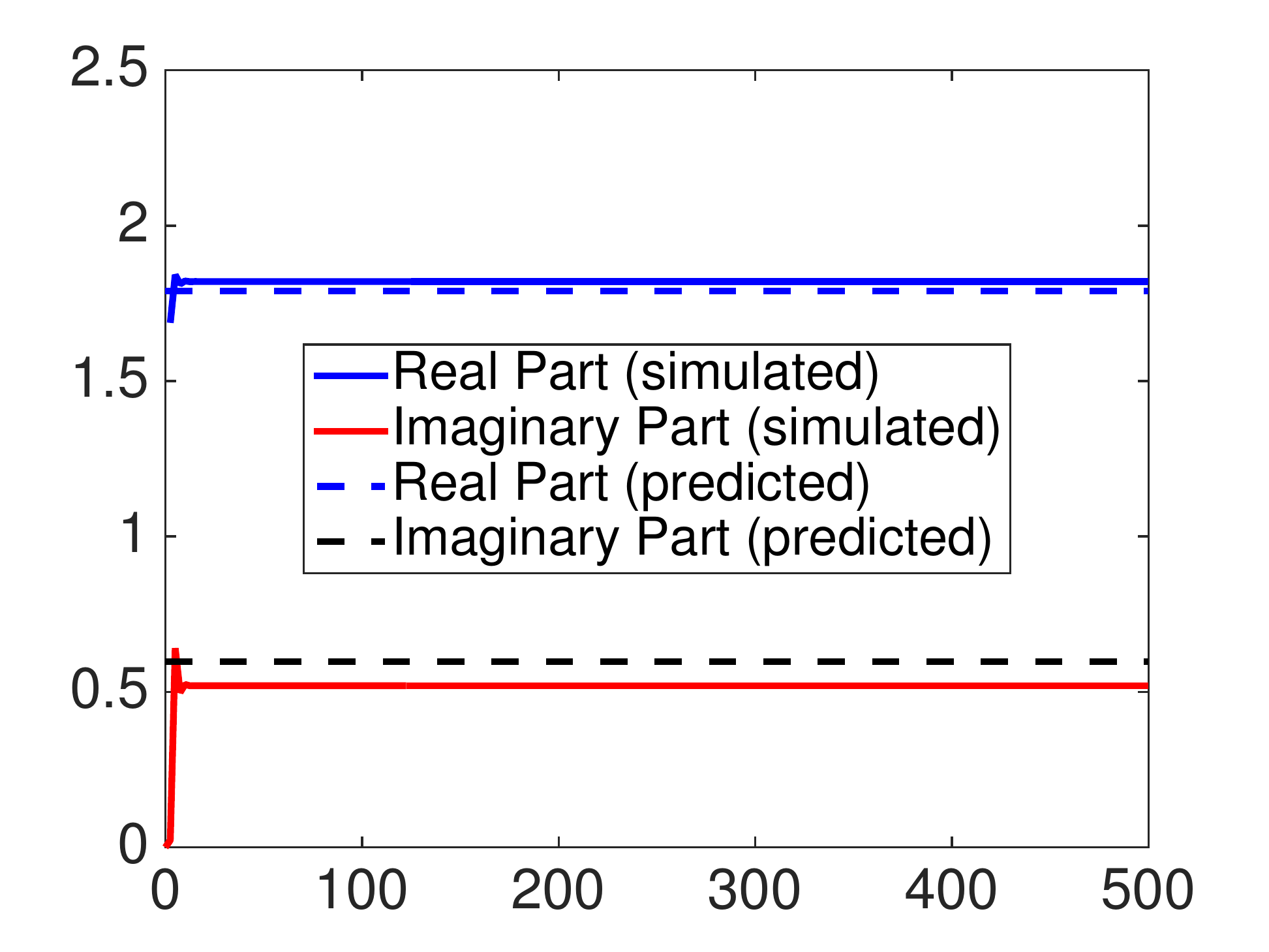}
  %\caption{$\epsilon=1$}
\put(19,13){(b) $\epsilon=1$}
\end{overpic}
%  \label{fig:sfig2}
\end{subfigure}
\caption{{Evolution of the real and imaginary parts of a
    ($N=3$) $\dn$ solution of \eqref{eq:lle} evaluated at $x=0$,
    with $G=0$, $\alpha = 1.7793$ (so that the initial waveform has $k^2 = 1-10^{-16}$), and $\epsilon$ set to} (a) $\epsilon=0.1$ and (b) $\epsilon=1$. The dotted lines are the theoretically calculated real and imaginary parts that result from our perturbation theory, whereas the solid lines are from the direct LLE simulation.  The perturbation theory holds remarkably well even at 
$\epsilon =1$.}
\label{fig:dn_evolve3}
\end{figure}

\section{Numerical Simulations}

{In this section, we compare
  numerical simulations
  of \eqref{eq:lle} with predictions made by the
  theory outlined above. In all
  simulations the value of $\beta$ is fixed, with $\beta = -0.01$.
  When $\epsilon \neq 0$, we set
  $F=(\rho(1+(\rho-\alpha)^2))^{1/2}$ with $\rho=0.95$ to remain
  in the right parameter space for the generation of
  frequency combs. The initial waveforms ($u$ at time zero)
  are set according to \eqref{eq:dn} with $\xi=\sigma=\sigma_0=x_0=0$
  and the value of $k$ determined by the detuning $\alpha$
  as in \eqref{eq:alpha}.}
First, we simulate the LLE to show that the
$\dn$ solution is stable, as opposed to the
observed instability of the $\cn$ and $\sn$ solutions
in Figure~\ref{fig:cn_sn}.  Figure~\ref{fig:dn_evolve1} shows
the evolution of the $\dn$ solution for $\epsilon=0.1$
{and the detuning $\alpha$ chosen so that} $k^2=0.9$ and
$k^2=1-10^{-12}\approx 1$. Recall
that for $k^2=0.9$ the linear stability analysis
showed strong instability and for $k^2=1-10^{-12}$
the analysis showed weaker instability, see
Figure~\ref{fig:dn_spec}. Further, recall that for both
values of the parameter, the solution should be unstable
for generic perturbations of the equation.
{In both simulations, the initial
  waveforms are corrupted} with
white noise in order to induce instability if it exists.
For $k^2 = 1-10^{-12}$, the pumping and damping terms of
the LLE, i.e. the LLE-specific perturbations,
have stabilized the $\dn$ solution. The $k^2 = 0.9$
solution is still unstable with this perturbation.
{In Fig.~\ref{fig:dn_evolve0}, we
  repeat these calculations without the LLE perturbation,
  i.e. setting $\epsilon=0$. The $k^2=1-10^{-12}$ solution
  is seen to be less stable than that in Fig.~\ref{fig:dn_evolve1}.

  In Fig.~\ref{fig:dn_evolve2},
  we plot equilibrated solutions of \eqref{eq:lle}
  as $\epsilon$ is increased.
  In this example, the ($N=3$) $\dn$-type solutions
  for $k^2=1-10^{-16}$ remain stable, even
  for large values of $\epsilon$. Note that the
  solutions deform away from the original $\dn$
  waveform and develop a pedestal as $\epsilon$ is increased.}
%The steady state solution is also deformed from the Jacobi elliptic function.
{Finally, Fig.~\ref{fig:dn_evolve3} contains plots of
the predicted time-independent} phase, determined by
\eqref{eq:phase_predict},
and the phase of a simulated microresonator solution
with a $\dn$ initial waveform, showing good agreement
between theory and simulation.

\begin{figure}[t]
\includegraphics[width=\linewidth]{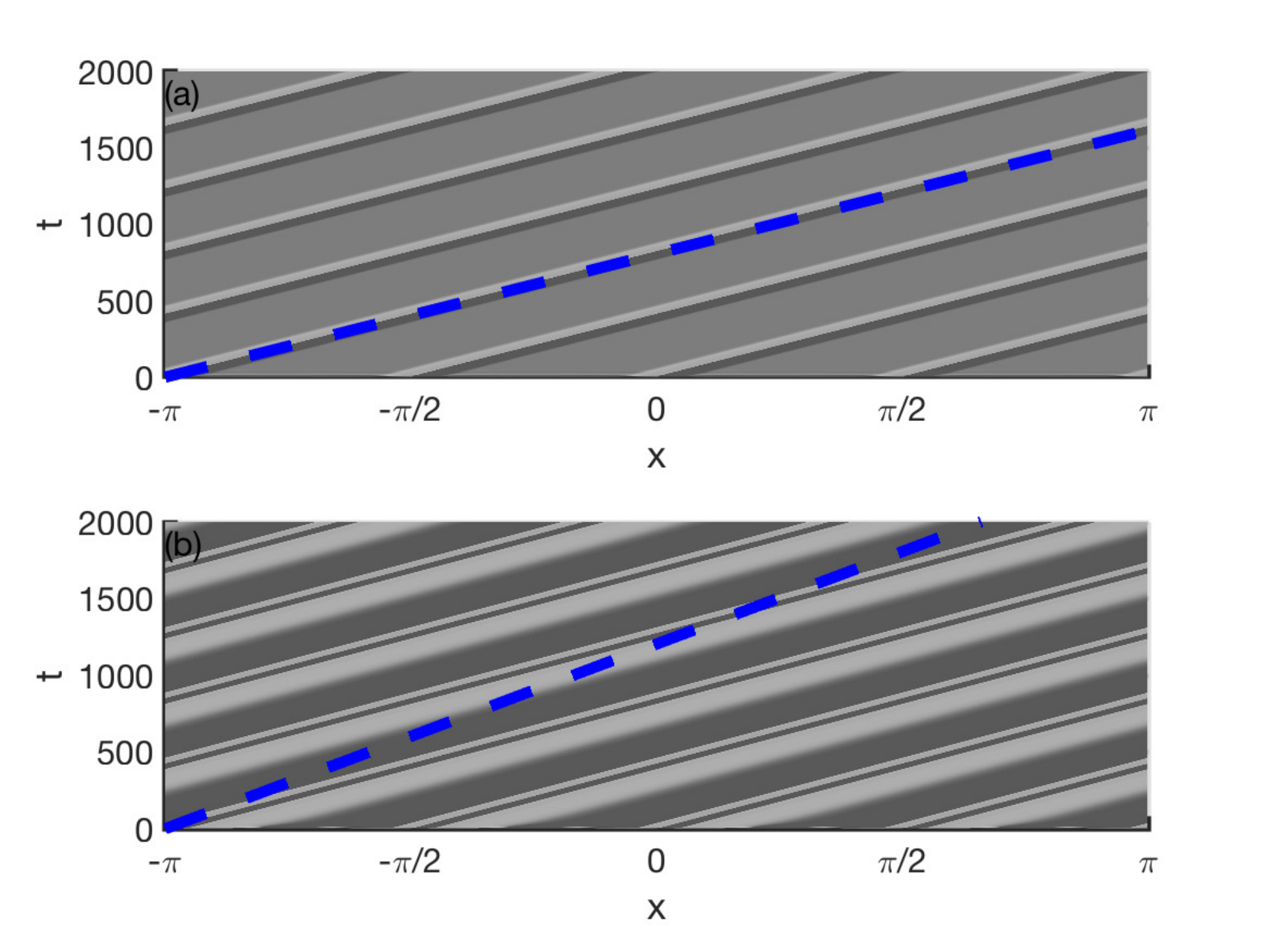}
\caption{{ Top view of a numerical simulation of \eqref{eq:lle} with $\alpha=1.8732$ and the addition of the Raman effect of \eqref{eq:GRaman}. The perturbation parameter $\epsilon$ is set to (a) $\epsilon=0.1$ and (b) $\epsilon=1$.  As predicted and quantified by our perturbation theory, the $\dn$ solution remains stable despite the induced drift of the solution. The drift velocity is compared with that computed from our theory using self-frequency shift in both cases. The dotted lines represent the theoretically calculated trajectories of the drift of the solutions. The perturbation theory holds well when $\epsilon$ is small.}}
\label{fig:Raman}
\end{figure}

\subsection{Raman term}

An important modification to the LLE equation is the addition of the Raman effect which is known to induce a self-frequency shift in the microresonator~\cite{karpov2016raman,milian2015solitons}.  The Raman effect is included in the LLE as part of the perturbation term $G(u,x,t)$ in \eqref{eq:lle}.  {Letting $U$ denote the waveform and $\mathfrak{G}(U)$ denote the Raman perturbation in physical units, we have}
\cite{karpov2016raman}
\begin{equation}
{\mathfrak{G}(U) = i \left[-f_R|U|^2 + f_{R}h_{R}\bigotimes |U|^{2} \right] U \approx -i \left[ f_{R}\tau_{R} \frac{\partial |U|^{2}}{\partial x} \right] U},
\label{eq:GRaman}
\end{equation}
where the constants $f_{R}$ and $\tau_{R}$ are the Raman fraction and the Raman shock time, respectively, and $\bigotimes$ denotes a convolution.  In simulations, the Raman response function $h_{R}$ is typically chosen to be \cite{blow1989theoretical}
\begin{equation}
h_{R}(x)=\frac{\tau_{1}^{2}+\tau_{2}^{2}}{\tau_{1}\tau_{2}^{2}}e^{-x/\tau_{2}}\sin(x/\tau_{1}),
\end{equation}
where $\tau_{1}=12.2$fs and $\tau_{2}=32$fs. {In our numerical simulation of the dimensionless LLE, \eqref{eq:lle}, the Raman term becomes $G(u)=-iC \frac{\partial |u|^{2}}{\partial x}u$, where $C=0.001$.}

The effect of the Raman perturbation of \eqref{eq:GRaman} can be 
substituted into the modulation constraints of \eqref{eq:pert2} to 
evaluate the impact on the comb dynamics. The symmetry properties of 
the perturbation play a large role in determining the resulting behavior.  
Specifically, symmetry considerations yield
\begin{subeqnarray}
&&  \frac{dx_{0}}{d\tau}=0 \; ,\\
&& \frac{d\sigma_{0}}{d\tau}=0 \; ,
\end{subeqnarray}
with the additional constraints that
\begin{subeqnarray}
&& \hspace*{-.3in} \frac{dB}{d\tau}=\frac{F\pi \cos(\sigma-\sigma_{0})}{\sqrt{-\beta}}-2B \; ,\\
&& \hspace*{-.3in} \frac{d\xi}{d\tau}=\frac{\langle2{C}BA^{3}k^{2}\dn^{2}z\sn z \cn z,\sn z\cn z\rangle}{\sqrt{-\beta}}   \neq 0 \, \slabel{eq:ramanmod}
\end{subeqnarray}
This determines the self-frequency shift induced by the 
Raman term since the value of $\xi$ gives the shift from 
the center frequency used to derive the LLE. 
In addition to the self-frequency shift, it should be recalled that 
\begin{equation}
\frac{dx_{0}}{dt}=-\beta \xi \, .
\end{equation}
As the term $\xi$ is slowly evolving, it can be thought of as
a constant over short time intervals so that the self-frequency 
shift generates a linear translation of the solution with a group 
velocity determined by the Raman term.  Importantly, 
the Raman term {\em does not} destabilize the comb, rather it simply 
shifts it in frequency and forces a translation.

{
  In Fig.~\ref{fig:Raman}, we plot simulations of the
  LLE with the addition of the Raman effect, i.e.
  $G(u)=-iC \frac{\partial |u|^{2}}{\partial x}u$,
  for both $\epsilon=0.1$ and $\epsilon=1$.
  The comb quickly forms and the induced translation
  is readily apparent. We also plot a line corresponding
  to the predicted drift velocity $dx_{0}/dt = -\beta \xi$.
  As noted above, only integer values of $\xi$ are allowed
  by the model. Nonetheless, the frequency shift that
  $\xi$ represents can be estimated from the simulation,
  and need not be integer valued. In particular, we take
  the empirical value of $\xi$ to be the center of mass
  of the Fourier coefficients of the simulated waveform
  (computed using the FFT). After the first few time steps,
  this value holds steady at approximately $\xi = 0.3890$ for
  the $\epsilon= 0.1$ simulation and $\xi = 0.2608$ for the
  $\epsilon = 1$ simulation. 
  The theoretical drift velocity matches well with the
  observed drift velocity of the simulation when
  $\epsilon=0.1$, whereas, for $\epsilon=1$, the prediction
  is not quantitatively satisfactory but corresponds to
  the qualitative behavior of the simulation (note that
  $\epsilon=1$ is far from the asymptotic regime).
}

\begin{figure}[t]
\includegraphics[width=\linewidth]{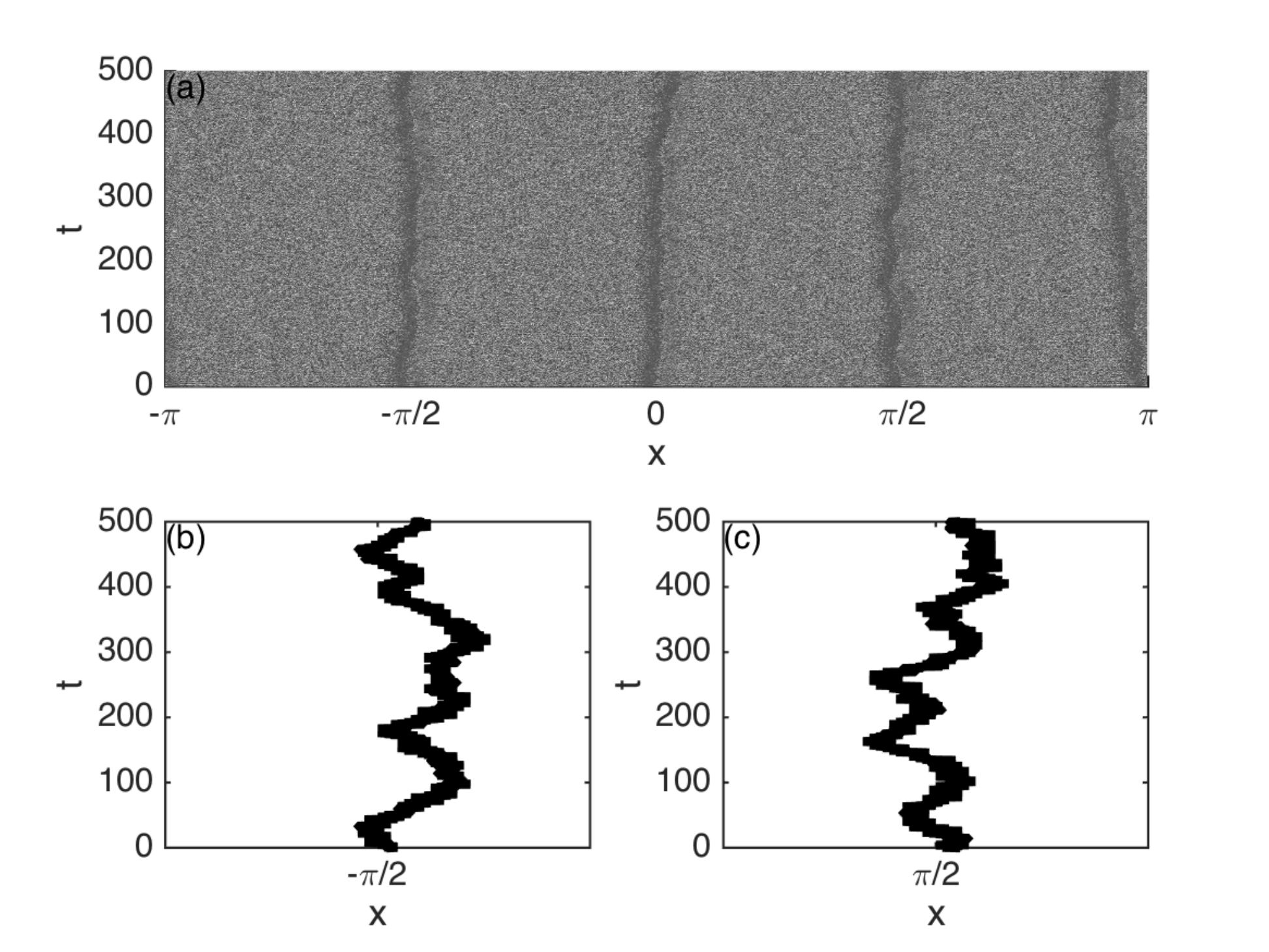}
\caption{(a) Top view of a numerical simulation of { \eqref{eq:lle} with $\epsilon = 0.1$, $\alpha=1.8732$, and} the addition of spontaneous emission noise { as defined in} \eqref{eq:wn}.  As predicted and quantified by our perturbation theory, the $\dn$ solution remains stable despite the induced random walk (drift) of the individual {pulse} solutions.  Much like the Gordon-Haus jitter, our perturbation theory captures the effect of the timing variance of individual pulses.  To highlight the random walk of each pulse, panels (b) and (c) show a detail of the pulses near $x=-\pi/2$ and $x=\pi/2$ respectively. }
\label{fig:Noise}
\end{figure}

\subsection{Spontaneous emission noise}

Spontaneous emission noise from pumping/amplification has always been a significant source of performance limitations in optical systems.  For instance, in optical communication systems, the noise from amplification results in the Gordon-Haus timing jitter~\cite{gordon1986random} which imposes a fundamental limit on transmission lengths for a given bit-error-rate constraint in lightwave communication systems.  Soliton perturbation theory provided the fundamental calculation of this limitation.  It also provided a number of engineering design strategies for trying to overcome the Gordon-Haus limitations, including sliding filters~\cite{mecozzi1991soliton,mollenauer1992sliding} and dispersion management~\cite{suzuki1995reduction,smith1996reduced,kutz1998gordon}.

The LLE perturbation theory developed here can also be used to evaluate the effects of spontaneous emission noise in the microresonator, something that has only recently been studied experimentally~\cite{herr2012universal,liao2017dependence}. Specifically, for this case the perturbation in \eqref{eq:lle} takes the form 
\begin{equation}
  G(u,x,t) = S(x,t) \; ,
  \label{eq:wn}
\end{equation}
where $S(x,t)$ is a white noise process modeling the spontaneous emission~\cite{weinstein1985modulational}.  In this case, for a specific realization of noise, the effects on the LLE comb parameters can be evaluated using \eqref{eq:pert2}.  Generically, the noise generates amplitude, phase, center-position and center-frequency jitter.  But the most pronounced effect comes from the fast scale dependency of the center position on the center frequency.  Thus the evolution
\begin{equation}
\frac{d\xi}{d\tau}=\frac{1}{\sqrt{-\beta}}(\langle\Re(e^{-i\psi} S(x,t) ),\sn z\cn z\rangle)
\end{equation}
produces a center frequency with mean $\langle \xi \rangle$ and variance $\langle \xi^2 \rangle$ which then
drives the center position through the relation $dx_0/dt = -\beta \xi$.  As with the Gordon-Haus jitter, this produces a jitter in the {pulse} position, leading to a degradation in performance.  Figure~\ref{fig:Noise} provides a simulation of the LLE under the influence of white noise perturbations \eqref{eq:wn}.  Note that the comb is stable, with fluctuations induced in the various solution parameters.  Most notably, the zoom in of the individual pulses shows the random-walk generated as a result of the noise.  As with Gordon-Haus jitter, the statistics of this random walk could be evaluated with the LLE perturbation theory we have developed.

\section{Conclusions}

In conclusion, we have shown that the LLE equation supports stable solutions of the Jacobi elliptic type.  These solutions model periodic pulse trains of soliton-like solutions for which the pumping $F$ is critical for stabilization.  Our rigorous stability analysis also results in a perturbation theory for characterizing the effects of higher-order terms in the microresonator, such as may arise from Raman scattering, higher-order dispersion and spontaneous emission noise.  The historical success of soliton perturbation theory in describing, for instance, Gordon-Haus timing jitter and/or the soliton self-frequency shifts, was critical in characterizing lightwave transmission systems and mode-locked lasers.  Similarly, the LLE perturbation theory presented here can be a critically enabling tool for characterizing a host of additional microresonator phenomenon and potentially engineering new resonator designs with improved performance metrics.

Our stability analysis helps confirm several experimental observations.  Most notably, it supports the recent observations that soliton states in the microresonator are not detuning degenerate, and can be individually addressed by laser detuning.  Indeed, the theory rigorously confirms that the detuning can be used to lock the microresonator to any target multiple-{pulse} state, where the stability of each multiple-{pulse} state is explicitly computed and its minimum detuning assessed.  The theory additionally shows that the phase-locking of the $\dn$ comb solution is an attractor to the resonator.  Moreover, only solutions with no nodal separation (a zero separating {pulses}) are stabilized.   Finally, the application of our theory to Raman scattering and stimulated emission perturbations show that neither effects destabilizes the comb.  Rather, they both generate a drift in the pulse train, one which is deterministic in nature (Raman) and one which produces a random walk (noise).

\section*{Acknowledgments}
J. N. Kutz acknowledges support from the Air Force
Office of Scientific Research (FA9550-17-1-0200).
The authors would also like to thank the anonymous
reviewers for their suggestions, which improved the
quality of the analysis.

\section*{Appendix}

In the following, we utilize some standard facts
concerning the eigenvalues and zeros of Sturm-Liouville
operators with periodic boundary conditions. See,
for example, Theorem 3.1 of Chapter 8 in \cite{coddington}.
These arguments are modeled after those in
\cite{weinstein1985modulational}.

\subsection{Proof of Proposition~\ref{prop:lplmnull}}

One can directly verify that $L_- [\dn z] = 0$.
Because $\dn z$ has no zeros, $\lambda = 0$ is the
first eigenvalue (listed in increasing order). 

Again, one can verify that $L_+ [\sn z \cn z] = 0$.
{ There is at most one function (up to
  a constant multiple) in $\ker (L_+)$ which is linearly
  independent of $w(z) = \sn z \cn z$. Note that the natural
  domain for $L_+$ is $H^2_\per[0,2NK)$ and recall that
    functions in $H^2_\per$ are determined by their values on
    $[0,2NK)$ and periodicity. For integer $j$, we have
      that $w(jK)=0$ and $w'(jK) = (-1)^j$.
      Suppose that $v$ is another solution
      of $L_+[v]=0$. We have that $w(z)v'(z) - w'(z)v(z)$
      is constant, so that $(v/w)' = d/w^2$ for some
      constant $d$ on any interval where $w \ne 0$.
      Consider an interval of the form $(jK,(j+1)K)$
      and let $x_j = (j+1/2)K$. For $jK<z<(j+1)K$, we have
    
  \begin{equation}
      v(z) = c_j w(z) + d_j w(z) \int_{x_j}^z \frac{dy}{w^2(y)} \; .
    \end{equation}
  Let

  \begin{equation}
    \tilde{w}_j(z) = w(z) \int_{x_j}^z \frac{dy}{w^2(y)}
  \end{equation}
  be defined on each interval $(jK,(j+1)K)$. 
  It can be verified that the limit of $\tilde{w}_j(z)$
  exists as you approach either endpoint. In particular,
  we have

  \begin{align*}
    \lim_{z\to 2jK^+} \tilde{w}_{2j}(z)
    &= \lim_{z\to 2jK^-} \tilde{w}_{2j-1}(z) =& -1\; , \\
    \lim_{z\to (2j+1)K^-} \tilde{w}_{2j}(z)
    &= \lim_{z\to (2j+1)K^+} \tilde{w}_{2j+1}(z) =& \frac{1}{\sqrt{1-k^2}} \; .
  \end{align*}
  Because $w$ is zero at all of these endpoints, we see
  that for $v$ to be continuous, the $d_j$
  should all be equal. Without loss of generality,
  we set $d_j = 1$ for all $j$.

  While the derivatives are still defined at the endpoints,
  they are not so well behaved. We have that

  \begin{align*}
    j_1 &:= \lim_{z\to 2jK^+} \tilde{w}'_{2j}(z)
    - \lim_{z\to 2jK^-} \tilde{w}'_{2j-1}(z) \\
    &= \frac{2}{1-k^2} \left ( (1-k)^{3/2} -1
    + (2-k^2) E(K/2,k) - (1-k^2) K \right ) \; ,\\
    j_2 &:= \lim_{z\to (2j+1)K^+} \tilde{w}'_{2j+1}(z) 
    - \lim_{z\to (2j+1)K^-} \tilde{w}'_{2j}(z) \\
    &= \sqrt{1-k^2} \left ( j_1 - \frac{2(2-k^2)E-4(1-k^2)K}{1-k^2}
    \right ) \; .
  \end{align*}
  Note that, for $0<k<1$, $j_1 \ne j_2$. To enforce
  that $v$ has continuous derivatives, we then
  obtain the following system of equations

  \begin{align*}
    c_1-c_0 &= -j_1 \\
    c_2-c_1 &= j_2 \\
    c_3-c_2 &= -j_1 \\
    &\vdots \\
    c_{2N-1}-c_{2N-2} &= -j_1 \\
    c_0-c_{2N-1} &= j_2 \; .
  \end{align*}
  By summing all of the equations, we obtain that
  $0 = N(j_2-j_1) \neq 0$, so that the equations are
  inconsistent. Therefore, there is no such $v$
  with a continuous derivative, i.e. there is no
  such $v$ in $H^2_\per[0,2NK)$. Note that for the case
    $k=0$, we see that $j_1 = j_2$ so that such a $v$
    does exist, as expected.
  }

\subsection{Proof of Proposition~\ref{prop:geneig}}

From Proposition~\ref{prop:lplmnull}, we have that

\begin{equation}
\ker({\bf L}^\dag) = \vecspan \{ (\dn z,0)^\intercal,
(0,\sn z \cn z)^\intercal\}  \; .
\end{equation}
Recall the definitions of $\phi$ and $\varphi$:
\begin{subeqnarray}
  && \hspace*{-.5in} \phi(z)=\left (K { E(z,k)}- E z \right)
  \dn z-k^{2}\sn z\cn z,\\
  && \hspace*{-.5in} \varphi(z)=k^{2} \cn z \sn z
  (K{ E(z,k)}-Ez)+(E-K) \dn z \nonumber\\
  && \hspace*{-.5in} \qquad \qquad+k^{2}K\cn^{2} z\dn z.
\end{subeqnarray}
It can be verified that

\begin{subeqnarray}
  && \hspace*{-.5in} L_{+}  L_{-}[\phi(z)]
  = L_{+} \left [-2k^{2}E\sn z\cn z\right] = 0,\\
  && \hspace*{-.5in} L_{-}  L_{+}[\varphi(z)]
  = L_{-} [2((k^{2} - 2)E-2(k^{2}-1)K)\dn z]=0.
\end{subeqnarray}
Therefore,
\begin{align}
  \ker( ({\bf L}^\dag)^2) &= \vecspan \{ (\dn z, 0)^\intercal,
  (0,\sn z \cn z)^\intercal, (\phi(z),0)^\intercal, \nonumber \\
  & \qquad (0,\varphi(z))^\intercal \} \; .
\end{align}
Suppose that $(f,g)^\intercal \in \ker( ({\bf L}^\dag)^3 )$.
Then, formally,

\begin{align}
  f &= c_1 L_-^{-1} \varphi(z)
  + c_2 \phi(z) + c_3 \dn z \; ,\\
  g &= c_4 L_+^{-1} \phi(z)
  + c_5 \varphi(z) + c_6 \sn z \cn z \; ,
\end{align}
where the inverses above denote a particular solution
of the corresponding inhomogeneous ODE. Consider
$L_-^{-1} \varphi(z)$.
Note that the Fredholm alternative implies that

\begin{align}
  0 &= \langle k^{2} \cn z \sn z (K{ E(z,k)}-Ez)+(E-K) \dn z, \dn z \rangle
  \nonumber \\ &\qquad + \langle k^{2}K\cn^{2} z\dn z, \dn z \rangle \\
  &= {N}(E^2 + (k^2-1)K^2) \; .
\end{align}
For $0<k<1$, the expression $E^2 + (k^2-1)K^2 > 0$, a contradiction.
Therefore, there is no such
particular solution. Similarly, consider
$L_+^{-1} \phi(z)$. The Fredholm alternative
implies that 

\begin{align}
  0 &= \langle \left (K{ E(z,k)}- Ez \right)
  \dn z-k^{2}K\sn z\cn z , \sn z \cn z \rangle \\
  &= -\frac{{N}}{k^2} \left (E^2 + (k^2-1)K^2 \right ) \; ,
\end{align}
again, a contradiction. Therefore,

\begin{equation}
\ker_g( {\bf L}^\dag ) = \ker( ({\bf L}^\dag)^2 ) \; .
\end{equation}

\subsection{Proof of Proposition~\ref{prop:h1control}}

The existence of $C_2$ is simple to establish. To
establish the existence of $C_1$, we require the
following two lemmas. {Note that for the remainder
of these statements, we assume that $N=1$.}

\begin{lemma} \label{lem:lpbound}
  Suppose that $\langle f, \dn z \rangle = 0$ and
  $\langle f, \phi(z) \rangle$. Then there exists
  a positive constant $C_1^+$ such that
  \begin{equation}
    \langle L_+ f, f \rangle \geq C_1^+ \| f\|^2_{L^2} \; .
  \end{equation}
\end{lemma}

\begin{lemma} \label{lem:lmbound}
  Suppose that $\langle g, \sn z \cn z \rangle = 0$ and
  $\langle g, \varphi(z) \rangle = 0$. Then
  there exists a positive constant $C_1^-$ such that
  \begin{equation}
    \langle L_- g, g \rangle \geq C_1^- \| g\|^2_{L^2} \; .
  \end{equation}  
\end{lemma}

Suppose that $\langle f, \dn z \rangle = 0$,
$\langle f, \phi(z) \rangle$,
$\langle g, \sn z \cn z \rangle = 0$, and
$\langle g, \varphi(z) \rangle = 0$.
Let $C_1^+$ and $C_1^-$ be as in Lemmas \ref{lem:lpbound}
and \ref{lem:lmbound}, respectively. Then

\begin{align}
  & \langle L_+ f,f \rangle + 6 \|f \|_{L^2}
  + \langle L_- g,g \rangle + 2 \| g \|_{L^2} \\
  & \quad = \| \dfrac{d}{dz} f \|_{L^2}
  + 6 \langle (1-\dn^2 z)f,f \rangle
  + (2-k^2) \|f \|_{L^2} \\
  & \quad \qquad + \| \dfrac{d}{dz} g \|_{L^2}
  + 2 \langle (1-\dn^2 z)g,g \rangle
  + (2-k^2) \|g \|_{L^2} \; , \\
  & \quad \geq \| f\|_{H^1_\per} + \| g \|_{H^1_\per} \; .
\end{align}
Therefore, the proposition holds with

\begin{equation}
  C_1 = \min \left ( \frac{1}{1+\frac{6}{C_1^+}},
  \frac{1}{1+\frac{2}{C_1^-}} \right ) \; .
\end{equation}

\subsubsection{Proof of Lemma \ref{lem:lpbound}}

In the following, we repeat the argument
of \cite{weinstein1985modulational}, making appropriate
changes to handle the periodic case. First,
we note that by Theorem 3.1 of Chapter 8 in
\cite{coddington}, $L_+$ has one negative
eigenvalue {(when $N=1$)}
with a corresponding eigenfunction
$f_0$, which we take to be nonnegative without
loss of generality. Define

\begin{equation}
  \gamma_1 = \min_f \langle L_+ f, f \rangle
  \; , \mbox{ where } \|f \|_2 = 1, \;
  \langle f, \dn z \rangle = 0 \; .
\end{equation}
Then, by Lemma E.1 of
\cite{weinstein1985modulational}, we have that
$\gamma_1 \geq 0$ if

\begin{equation}
  \langle L_+^{-1} \dn z, \dn z \rangle
  \leq 0 \; ,
\end{equation}
which is straightforward to verify using arguments
similar to those in the proof of Proposition~\ref{prop:geneig}.
Therefore,
$\gamma_1 \geq 0$. The lemma is then proved if we can
show that $ \gamma_2 = \inf_{f} \langle L_+ f, f \rangle$
with $f$ restricted such that $\|f\|_{L^2} = 1$,
$\langle f, \dn z \rangle = 0$,
and $\langle f, \phi(z) \rangle = 0$ is non-zero,
as $\gamma_2 \geq \gamma_1 \geq 0$.

Suppose that $\gamma_2 = 0$. Let $f_m$ be a minimizing
sequence of $\langle L_+ f, f \rangle$ satisfying
$\|f_m\|_{L^2} = 1$, $\langle f_m, \dn z\rangle = 0$,
and $\langle f_m, \phi(z) \rangle = 0$. Given $\delta > 0$,
there exists a $M(\delta)$ such that

\begin{align}
  0 &< \int_{-K(k)}^{K(k)} \left (\frac{d}{dz} f_m \right )^2
  \, dz
  + (2-k^2) \int_{-K(k)}^{K(k)} f_m^2 \, dz \\
  & \leq 6 \int_{-K(k)}^{K(k)} \dn^2 z f_m^2 \, dz + \delta \; ,
\end{align}
for all $m \geq M(\delta)$. In particular, the sequence
$f_m$ is uniformly bounded in the $H^1_\per$ norm.
Therefore, there is a subsequence of $f_m$ which converges weakly
to an $H^1_\per$ function $f_*$. This function
satisfies the constraints $\langle f_*, \dn z \rangle = 0$ and
$\langle f_*, \phi(z) \rangle = 0$ by weak convergence.
Because $H^1_\per$ is compactly
embedded in $L^2$, there exists a further subsequence, which
we denote by $f_{m_j}$, that converges in the $L^2$ norm
(to $f_*$). Therefore, $ \| f_* \|_{L^2} = 1$. 

Let $h$ be such that $\|h\|_2 = 1$. Note that 
$\langle h, f_*' \rangle = \lim \langle h, f_{m_j}' \rangle
\leq \liminf \| f_{m_j}' \|_{L^2}$ by weak convergence
in $H^1_\per$. 
Taking the maximum over all such $h$ implies that 
$\| f_*' \|_{L^2} \leq \liminf \| f_{m_j}' \|_{L^2}$.
Combining this with the $L^2$ convergence of $f_{m_j}$
gives that $\langle L_+ f_*, f_* \rangle \leq \liminf
\langle L_+ f_{m_j}, f_{m_j} \rangle = 0$, so that 
$\langle L_+ f_*, f_* \rangle = 0$. 

Because $f_*$ attains the minimum and is admissible,
there exists a critical point of the problem

\begin{align}
  (L_+ - \lambda_1)f &= \lambda_2 \dn z + \lambda_3 \phi(z)
  \; , \label{eq:lagmul1} \\
  \|f \|_2 &= 1 \; , \\
  \langle f, \dn z \rangle &= 0 \; ,\\
  \langle f, \phi(z) \rangle &= 0 \; ,
\end{align}
of the form $(f_*, \lambda_1, \lambda_2, \lambda_3)$.
Taking the inner product of $f_*$ with \eqref{eq:lagmul1},
we obtain that $\lambda_1 = \langle L_+f_*,f_*\rangle = 0$.
This implies that

\begin{equation}
  L_+ f_* = \lambda_2 \dn z + \lambda_3 \phi(z) \; .
  \label{eq:lagmul2}
\end{equation}
Taking the inner product of $\sn z \cn z$ with
\eqref{eq:lagmul2}, we obtain that $\lambda_3 = 0$.
Following the arguments in the proof of
Proposition~\ref{prop:geneig}, this implies that

\begin{equation}
  f_* = \frac{\lambda_2}{2((k^2-2)E-2(k^2-1)K)} \varphi(z)
  + \lambda_4 \sn z \cn z \; ,
\end{equation}
for some $\lambda_4$. The constraint $\langle f_*, \phi(z) \rangle = 0$
implies that $\lambda_4 = 0$ and the constraint
$\langle f_*, \dn z \rangle = 0$ implies that $\lambda_2 = 0$.
We have that $f_* \equiv 0$, a contradiction.
Therefore, $\gamma_2 > 0$, proving the lemma.

\subsubsection{Proof of Lemma \ref{lem:lmbound}}

This lemma can be proved using arguments similar
to the above.

% Bibliography
\bibliography{refs}

\begin{thebibliography}{10}
\newcommand{\enquote}[1]{``#1''}

\bibitem{app2}
S.~T. Cundiff and A.~M. Weiner, \enquote{Optical arbitrary waveform
  generation,} Nature Photonics \textbf{4}, 760--766 (2010).

\bibitem{kippenberg2011microresonator}
T.~J. Kippenberg, R.~Holzwarth, and S.~Diddams, \enquote{Microresonator-based
  optical frequency combs,} Science \textbf{332}, 555--559 (2011).

\bibitem{del2011octave}
P.~DelHaye, T.~Herr, E.~Gavartin, M.~Gorodetsky, R.~Holzwarth, and T.~J.
  Kippenberg, \enquote{Octave spanning tunable frequency comb from a
  microresonator,} Physical Review Letters \textbf{107}, 063901 (2011).

\bibitem{papp2011spectral}
S.~B. Papp and S.~A. Diddams, \enquote{Spectral and temporal characterization
  of a fused-quartz-microresonator optical frequency comb,} Physical Review A
  \textbf{84}, 053833 (2011).

\bibitem{ferdous2011spectral}
F.~Ferdous, H.~Miao, D.~E. Leaird, K.~Srinivasan, J.~Wang, L.~Chen, L.~T.
  Varghese, and A.~M. Weiner, \enquote{Spectral line-by-line pulse shaping of
  on-chip microresonator frequency combs,} Nature Photonics \textbf{5},
  770--776 (2011).

\bibitem{app1}
J.~Pfeifle, V.~Brasch, M.~Lauermann, Y.~Yu, D.~Wegner, T.~Herr, K.~Hartinger,
  P.~Schindler, J.~Li, D.~Hillerkuss \emph{et~al.}, \enquote{Coherent terabit
  communications with microresonator kerr frequency combs,} Nature photonics
  \textbf{8}, 375--380 (2014).

\bibitem{liang2015high}
W.~Liang, D.~Eliyahu, V.~Ilchenko, A.~Savchenkov, A.~Matsko, D.~Seidel, and
  L.~Maleki, \enquote{High spectral purity kerr frequency comb radio frequency
  photonic oscillator,} Nature communications \textbf{6} (2015).

\bibitem{suh2016microresonator}
M.-G. Suh, Q.-F. Yang, K.~Y. Yang, X.~Yi, and K.~J. Vahala,
  \enquote{Microresonator soliton dual-comb spectroscopy,} Science
  \textbf{354}, 600--603 (2016).

\bibitem{herr2014temporal}
T.~Herr, V.~Brasch, J.~Jost, C.~Wang, N.~Kondratiev, M.~Gorodetsky, and
  T.~Kippenberg, \enquote{Temporal solitons in optical microresonators,} Nature
  Photonics \textbf{8}, 145--152 (2014).

\bibitem{yi2015soliton}
X.~Yi, Q.-F. Yang, K.~Y. Yang, M.-G. Suh, and K.~Vahala, \enquote{Soliton
  frequency comb at microwave rates in a high-q silica microresonator,} Optica
  \textbf{2}, 1078--1085 (2015).

\bibitem{kutz1}
B.~G. Bale, K.~Kieu, J.~N. Kutz, and F.~Wise, \enquote{Transition dynamics for
  multi-pulsing in mode-locked lasers,} Opt. Express \textbf{17}, 23137--23146
  (2009).

\bibitem{kutz2}
M.~O. Williams, E.~Shlizerman, and J.~N. Kutz, \enquote{The multi-pulsing
  transition in mode-locked lasers: a low-dimensional approach using waveguide
  arrays,} J. Opt. Soc. Am. B \textbf{27}, 2471--2481 (2010).

\bibitem{kutz3}
F.~Li, P.~K.~A. Wai, and J.~N. Kutz, \enquote{Geometrical description of the
  onset of multi-pulsing in mode-locked laser cavities,} J. Opt. Soc. Am. B
  \textbf{27}, 2068--2077 (2010).

\bibitem{kip1}
T.~Herr, V.~Brasch, J.~Jost, C.~Wang, N.~Kondratiev, M.~Gorodetsky, and
  T.~Kippenberg, \enquote{Temporal solitons in optical microresonators,} Nature
  Photonics \textbf{8}, 145--152 (2014).

\bibitem{kip2}
H.~Guo, M.~Karpov, E.~Lucas, A.~Kordts, M.~H. Pfeiffer, V.~Brasch, G.~Lihachev,
  V.~E. Lobanov, M.~L. Gorodetsky, and T.~J. Kippenberg, \enquote{Universal
  dynamics and deterministic switching of dissipative kerr solitons in optical
  microresonators,} Nature Physics  (2016).

\bibitem{lugiato1987spatial}
L.~A. Lugiato and R.~Lefever, \enquote{Spatial dissipative structures in
  passive optical systems,} Physical review letters \textbf{58}, 2209 (1987).

\bibitem{PhysRevA.87.053852}
Y.~K. Chembo and C.~R. Menyuk, \enquote{Spatiotemporal lugiato-lefever
  formalism for kerr-comb generation in whispering-gallery-mode resonators,}
  Phys. Rev. A \textbf{87}, 053852 (2013).

\bibitem{karpman1977perturbation}
V.~Karpman and E.~Maslov, \enquote{Perturbation theory for solitons,} JETP
  \textbf{73}, 537--559 (1977).

\bibitem{kodama1981perturbations}
Y.~Kodama and M.~J. Ablowitz, \enquote{Perturbations of solitons and solitary
  waves,} Studies in Applied Mathematics \textbf{64}, 225--245 (1981).

\bibitem{kaup1990perturbation}
D.~Kaup, \enquote{Perturbation theory for solitons in optical fibers,} Physical
  Review A \textbf{42}, 5689 (1990).

\bibitem{elgin1993perturbations}
J.~Elgin, \enquote{Perturbations of optical solitons,} Physical Review A
  \textbf{47}, 4331 (1993).

\bibitem{kartner1996soliton}
F.~Kartner, I.~Jung, and U.~Keller, \enquote{Soliton mode-locking with
  saturable absorbers,} IEEE Journal of Selected Topics in Quantum Electronics
  \textbf{2}, 540--556 (1996).

\bibitem{kapitula2002stability}
T.~Kapitula, J.~N. Kutz, and B.~Sandstede, \enquote{Stability of pulses in the
  master mode-locking equation,} JOSA B \textbf{19}, 740--746 (2002).

\bibitem{kapitula2004evans}
T.~Kapitula, N.~Kutz, and B.~Sandstede, \enquote{The evans function for
  nonlocal equations,} Indiana University mathematics journal pp. 1095--1126
  (2004).

\bibitem{Bale:08}
B.~G. Bale and J.~N. Kutz, \enquote{Variational method for mode-locked lasers,}
  J. Opt. Soc. Am. B \textbf{25}, 1193--1202 (2008).

\bibitem{PhysRevA.89.063814}
C.~Godey, I.~V. Balakireva, A.~Coillet, and Y.~K. Chembo, \enquote{Stability
  analysis of the spatiotemporal lugiato-lefever model for kerr optical
  frequency combs in the anomalous and normal dispersion regimes,} Phys. Rev. A
  \textbf{89}, 063814 (2014).

\bibitem{cross1993pattern}
M.~C. Cross and P.~C. Hohenberg, \enquote{Pattern formation outside of
  equilibrium,} Reviews of modern physics \textbf{65}, 851 (1993).

\bibitem{carr2000stationary}
L.~D. Carr, C.~W. Clark, and W.~P. Reinhardt, \enquote{Stationary solutions of
  the one-dimensional nonlinear schr{\"o}dinger equation. ii. case of
  attractive nonlinearity,} Physical Review A \textbf{62}, 063611 (2000).

\bibitem{bronski2001bose}
J.~C. Bronski, L.~D. Carr, B.~Deconinck, and J.~N. Kutz, \enquote{Bose-einstein
  condensates in standing waves: The cubic nonlinear schr{\"o}dinger equation
  with a periodic potential,} Physical Review Letters \textbf{86}, 1402 (2001).

\bibitem{bronski1}
J.~C. Bronski, L.~D. Carr, B.~Deconinck, J.~N. Kutz, and K.~Promislow,
  \enquote{Stability of repulsive bose-einstein condensates in a periodic
  potential,} Physical Review E \textbf{63}, 036612 (2001).

\bibitem{bronski2}
J.~C. Bronski, L.~D. Carr, R.~Carretero-Gonz{\'a}lez, B.~Deconinck, J.~N. Kutz,
  and K.~Promislow, \enquote{Stability of attractive bose-einstein condensates
  in a periodic potential,} Physical Review E \textbf{64}, 056615 (2001).

\bibitem{bender2013advanced}
C.~M. Bender and S.~A. Orszag, \emph{Advanced mathematical methods for
  scientists and engineers I: Asymptotic methods and perturbation theory}
  (Springer Science \& Business Media, 2013).

\bibitem{kevorkian2013perturbation}
J.~Kevorkian and J.~D. Cole, \emph{Perturbation methods in applied
  mathematics}, vol.~34 (Springer Science \& Business Media, 2013).

\bibitem{weinstein1985modulational}
M.~I. Weinstein, \enquote{Modulational stability of ground states of nonlinear
  schr{\"o}dinger equations,} SIAM journal on mathematical analysis
  \textbf{16}, 472--491 (1985).

\bibitem{Bernard}
N.~Bottman, B.~Deconinck, and M.~Nivala, \enquote{Elliptic solutions of the
  defocusing nls equation are stable,} Journal of Physics A: Mathematical and
  Theoretical \textbf{44}, 285201 (2011).

\bibitem{deconinck2017stability}
B.~Deconinck and B.~L. Segal, \enquote{The stability spectrum for elliptic
  solutions to the focusing nls equation,} Physica D: Nonlinear Phenomena
  \textbf{346}, 1--19 (2017).

\bibitem{stabilityDn}
S.~Gustafson, S.~Le~Coz, and T.-P. Tsai, \enquote{Stability of periodic waves
  of 1d cubic nonlinear schr{\"o}dinger equations,} Applied Mathematics
  Research eXpress \textbf{2017}, 431--487 (2017).

\bibitem{karpov2016raman}
M.~Karpov, H.~Guo, A.~Kordts, V.~Brasch, M.~H. Pfeiffer, M.~Zervas,
  M.~Geiselmann, and T.~J. Kippenberg, \enquote{Raman self-frequency shift of
  dissipative kerr solitons in an optical microresonator,} Physical review
  letters \textbf{116}, 103902 (2016).

\bibitem{milian2015solitons}
C.~Mili{\'a}n, A.~V. Gorbach, M.~Taki, A.~V. Yulin, and D.~V. Skryabin,
  \enquote{Solitons and frequency combs in silica microring resonators:
  Interplay of the raman and higher-order dispersion effects,} Physical Review
  A \textbf{92}, 033851 (2015).

\bibitem{blow1989theoretical}
K.~J. Blow and D.~Wood, \enquote{Theoretical description of transient
  stimulated raman scattering in optical fibers,} IEEE Journal of Quantum
  Electronics \textbf{25}, 2665--2673 (1989).

\bibitem{gordon1986random}
J.~P. Gordon and H.~A. Haus, \enquote{Random walk of coherently amplified
  solitons in optical fiber transmission,} Optics letters \textbf{11}, 665--667
  (1986).

\bibitem{mecozzi1991soliton}
A.~Mecozzi, J.~D. Moores, H.~A. Haus, and Y.~Lai, \enquote{Soliton transmission
  control,} Optics letters \textbf{16}, 1841--1843 (1991).

\bibitem{mollenauer1992sliding}
L.~F. Mollenauer, J.~P. Gordon, and S.~G. Evangelides, \enquote{The
  sliding-frequency guiding filter: an improved form of soliton jitter
  control,} Optics letters \textbf{17}, 1575--1577 (1992).

\bibitem{suzuki1995reduction}
M.~Suzuki, I.~Morita, N.~Edagawa, S.~Yamamoto, H.~Taga, and S.~Akiba,
  \enquote{Reduction of gordon-haus timing jitter by periodic dispersion
  compensation in soliton transmission,} Electronics Letters \textbf{31},
  2027--2029 (1995).

\bibitem{smith1996reduced}
N.~Smith, W.~Forysiak, and N.~Doran, \enquote{Reduced gordon-haus jitter due to
  enhanced power solitons in strongly dispersion managed systems,} Electronics
  Letters \textbf{32}, 2085--2086 (1996).

\bibitem{kutz1998gordon}
J.~N. Kutz and P.~Wai, \enquote{Gordon-haus timing jitter reduction in
  dispersion-managed soliton communications,} IEEE Photonics Technology Letters
  \textbf{10}, 702--704 (1998).

\bibitem{herr2012universal}
T.~Herr, K.~Hartinger, J.~Riemensberger, C.~Wang, E.~Gavartin, R.~Holzwarth,
  M.~Gorodetsky, and T.~Kippenberg, \enquote{Universal formation dynamics and
  noise of kerr-frequency combs in microresonators,} Nature Photonics
  \textbf{6}, 480--487 (2012).

\bibitem{liao2017dependence}
P.~Liao, C.~Bao, A.~Kordts, M.~Karpov, M.~H. Pfeiffer, L.~Zhang,
  A.~Mohajerin-Ariaei, Y.~Cao, A.~Almaiman, M.~Ziyadi \emph{et~al.},
  \enquote{Dependence of a microresonator kerr frequency comb on the pump
  linewidth,} Optics Letters \textbf{42}, 779--782 (2017).

\bibitem{coddington}
E.~A. Coddington and N.~Levinson, \emph{Theory of ordinary differential
  equations} (Tata McGraw-Hill Education, 1955).

\end{thebibliography}

%Manual citation list
%\begin{thebibliography}{1}
%\bibitem{Zhang:14}
%Y.~Zhang, S.~Qiao, L.~Sun, Q.~W. Shi, W.~Huang, %L.~Li, and Z.~Yang,
 % \enquote{Photoinduced active terahertz metamaterials with nanostructured
  %vanadium dioxide film deposited by sol-gel method,} Opt. Express \textbf{22},
  %11070--11078 (2014).
%\end{thebibliography}

% Please include bios and photos of all authors for aop articles 

\end{document}